\documentclass[
 aps,prapplied,
 amsmath,amssymb,
 reprint,
]{revtex4-2}

\usepackage{graphicx}
\usepackage{dcolumn}
\usepackage{bm}

\usepackage{url}
\usepackage[colorlinks=true,linkcolor=blue,citecolor=blue,urlcolor=blue]{hyperref}

\usepackage[utf8]{inputenc}
\usepackage[T1]{fontenc}
\usepackage{mathptmx}
\usepackage{etoolbox}
\usepackage{xcolor}
\usepackage[normalem]{ulem}

\colorlet{RED}{red}

\begin{document}


\title{Hydrogen contamination induced device variability in Al/AlO$_x$/Al Josephson junctions}
\author{Yu Zhu}
\email[Corresponding author: ]{eric@nanoacademic.com}
\affiliation{Nanoacademic Technologies Inc., Suite 802, 666 rue Sherbrooke Ouest, Montréal, Québec H3A 1E7, Canada}

\author{Aldilene Saraiva-Souza}
\affiliation{Nanoacademic Technologies Inc., Suite 802, 666 rue Sherbrooke Ouest, Montréal, Québec H3A 1E7, Canada}

\author{F\'elix Beaudoin}
\affiliation{Nanoacademic Technologies Inc., Suite 802, 666 rue Sherbrooke Ouest, Montréal, Québec H3A 1E7, Canada}

\author{Hong Guo}
\affiliation{Department of Physics, McGill University, Montréal, Québec, H3A 2T8, Canada}

\date{\today}

\begin{abstract}
Hydrogen contamination in Josephson junctions is a potential source of
device-to-device variability in superconducting
qubits, but the incorporation, location, and bonding of hydrogen in the tunnel barrier remain poorly characterized. In this work, we investigate hydrogen incorporation in oxidized
aluminum barriers by combining molecular dynamics simulations with atomistic
quantum transport calculations. The oxide growth simulations are performed
by molecular dynamics for Al surfaces exposed to dense O$_{\text{2}}$ and H$_{\text{2}%
}$O environments, yielding amorphous AlO$_{\text{x}}$ layers with hydrogen
content comparable to experimentally relevant levels. From $400$
statistically independent samples, we find that the number of H atoms in the
oxide is well described by a beta-binomial distribution, reflecting
correlations induced by the self-limiting oxidation process. Structural
analysis shows that most hydrogen atoms reside near the AlO$_{\text{x}}$
surface and predominantly form Al-OH and Al-OH-Al motifs. To assess the
impact of hydrogen on transport, we construct Al/Al$_{\text{2}}$O$_{\text{3}}
$/Al junction models and perform quantum transport analysis. The H atoms are found to 
increase the transmission coefficient near the Fermi level by forming an interface dipole that shifts the oxide bands relative to the Fermi level of the electrodes. By combining the H atom
number statistics from molecular dynamics with the transmission coefficients
from quantum transport calculations, we predict a probability distribution
for the Josephson energy. For a Josephson junction with an average hydrogen
content of $2.56$ at.\%, the resulting Josephson energy is predicted to be $%
E_{J}/h=10.23$ GHz with a standard deviation of only $3.6$ MHz. Hydrogen therefore shifts the mean Josephson energy by about $11\%$, in proportion to the hydrogen content, while its contribution to device-to-device variability remains significantly smaller than the variability observed in state-of-the-art fabricated junctions. These results provide an atomistic picture of
hydrogen contamination in Josephson
junctions and of its effect on the Josephson energy.
\end{abstract}
\maketitle

\section{Introduction}

The superconducting quantum-computing landscape has undergone a
transformative shift in both scale and performance, highlighted by the
deployment of increasingly large processors such as IBM's 1,121-qubit Condor 
\cite{ref-condor}, Google's 105-qubit Willow \cite{ref-willow}, and
Rigetti's 84-qubit Ankaa-2 system \cite{ref-ankaa2}. However, continued
scaling of these architectures ultimately hinges on resolving fundamental
materials science challenges \cite{ref-materials-science-qc}. A prominent
issue is hydrogen contamination in Al/AlO$_{\text{x}}$/Al Josephson
junctions. Josephson junctions show device-to-device variability
\cite{Kreikebaum2020JJvariation,Osman2021Xmon,Sulangi2020JAP,Pappas2024ABAA},
to which such contamination may contribute. This variability is
often mitigated using frequency-tunable qubits; however, the
resulting flux-noise sensitivity \cite%
{Yoshihara2006FluxNoise,Bylander2011FluxNoise,Quintana2017FluxNoise,Hutchings2017FluxIndependent,Mergenthaler2021TLS}
and calibration overhead \cite{Barrett2023PRApp,Marciniak2026Hydrogen} may
become major practical obstacles as qubit counts scale by orders of
magnitude toward fault-tolerant quantum computing. Furthermore,
hydrogen-related defects can act as two-level systems (TLS) leading to
charge noise \cite%
{Gordon2014TLS,Holder2013TLS,Muller2019TLSReview,Lisenfeld2019Review}, or as
magnetic defects leading to flux noise \cite{Wang2018PRB}, which limits the
coherence time of the quantum state and overall qubit lifetimes. 
Despite growing experimental evidence for hydrogen-related losses, a quantitative first-principles understanding of how hydrogen incorporation modifies the electronic tunneling through the oxide barrier remains largely unexplored. 

The two primary methods used to fabricate Josephson junctions are shadow
evaporation (Dolan bridge) and Manhattan-style (overlap) evaporation; in both, the Al electrodes are deposited by electron-beam physical vapor deposition (EBPVD), i.e., e-beam evaporation.
Regardless of the method, the AlO$_{\text{x}}$ tunnel barrier is the most
sensitive region, because the junction's critical current $I_{c}$ (and hence
the qubit frequency) depends exponentially on the thickness and quality of
the tunnel barrier. Even when the oxide is formed in situ under high vacuum,
hydrogen can still be incorporated into AlO$_{\text{x}}$ through residual
adsorbed water and hydrocarbons, resist- or etch-related residues, and
subsequent processing or air exposure. In particular, secondary-ion mass spectrometry (SIMS) depth profiling of Al/AlO$_{\text{x}}$/Al trilayers reports peak hydrogen densities in the barrier of $1.37\times 10^{21}$~cm$^{-3}$ for an EBPVD-grown junction and $(4.18$--$4.86)\times 10^{21}$~cm$^{-3}$ for magnetron-sputtered junctions, corresponding to $1.1$ and $3.5$--$4.1$~at.$\%$ relative to the atomic density of Al$_2$O$_3$ \cite{ref-JJ-Aaron-thesis}. Neutron spectroscopy of ambient-exposed Al/Al$_2$O$_3$ nanoparticles has revealed a quasielastic signal attributed to the motion of hydrogen, and thermogravimetric analysis indicated about $0.3$ H$_2$O molecules per nm$^{2}$ bound to the amorphous alumina surface~\cite{ref-Cortie2020}. 
However, these experimental characterizations do not directly reveal how individual hydrogen defects alter the tunneling conductance, highlighting the need for complementary atomistic transport modeling. 

Device-to-device variability of Josephson junctions has many possible sources, including variations in the oxide thickness, barrier stoichiometry, junction area, interface roughness, pinholes, and contamination. In previous work, we studied the variability of $E_{J}$ caused by interface roughness using a continuum model of the barrier thickness~\cite{ref-Zhu2026-APL}. Hydrogen contamination is a different source: its effect on transport is determined by individual H atoms and their bonding environment, and therefore requires atomistic modeling. The focus of this research is to investigate hydrogen contamination in Al/AlO%
$_{\text{x}}$/Al Josephson junctions in the presence of H$_{\text{2}}$O
molecules. We aim to address three questions: (1) What is the probability
distribution for the number of H atoms among different Josephson junctions?
(2) Where are the H atoms physically located within the Al/AlO$_{\text{x}}$
interface? (3) What are the impacts of H atoms on the critical current and
Josephson energy? We do not address whether the identified H motifs act as TLS; they are candidates from which TLS may arise, and only a preliminary analysis of their potential-energy landscape is presented here. To address these questions, we combine molecular dynamics simulations using CHGNet, a pretrained universal machine-learning interatomic potential~\cite{ref-chgnet}, with nonequilibrium Green's function density functional theory (NEGF-DFT) calculations in three steps: (1) MD simulations give the distribution of the number of H atoms per MD cell and their binding sites (Secs.~II and III); (2) NEGF-DFT calculations give the change in the transmission coefficient caused by one H atom (Sec.~IV); and (3) a statistical model combines the two to give the distribution of $E_{J}$ for a junction of realistic area (Sec.~V). We found that the number of H atoms across
different Al/AlO$_{\text{x}}$ tunnel barriers is well described by a beta-binomial distribution. Across 400 Al/AlO$_{\text{x}}$ interfaces, 91.3\% of the H atoms form hydroxyl (-OH) groups bonded to one Al atom (terminal $\mu_{1}$-OH) or bridging two or, rarely, three Al atoms ($\mu_{2}$- and $\mu_{3}$-OH), 5.4\% H atoms bind to an Al atom through chemisorbed H2O-like motifs, while the remaining H atoms form a
variety of rare configurations. Combining these statistics with the transport calculations, we find that H atoms increase the mean Josephson energy in proportion to the hydrogen content, whereas the device-to-device variability caused by the random incorporation of hydrogen remains significantly smaller than the variability observed in state-of-the-art fabricated junctions.

The rest of the paper is organized as follows. In the next section, molecular dynamics is presented to determine the growth of AlO$_{\text{x}}$ on Al surface with some H$_2$O molecules. In Section III, structural analysis at the atomic level is performed to determine the distribution of H atoms. In Section IV, quantum transport calculation is presented in terms of the transmission coefficients at the Fermi level, which is used in Section V to predict the Josephson energy $E_J$ via the Ambegaokar-Baratoff relation \cite{ambegaokar1963tunneling}. Finally, Section VI is reserved for a short summary, and the dependence of the band gap and the transmission coefficients on the Hubbard parameter $U$ is presented in Appendix~\ref{app:U}.

\section{Molecular dynamics}

To address these questions, we perform molecular dynamics (MD) simulations
of AlO$_{\text{x}}$ growth on Al surfaces. In nearly all high-coherence
transmons, Josephson junctions are fabricated by double-angle shadow
evaporation followed by thermal O2 oxidation. The resulting AlO$_{\text{x}}$
barrier is amorphous, with a typical thickness of about 1-2 nm. Since the
oxidation is a self-limiting process, the tunnel barrier thickness is
relatively uniform across samples. In experiments, the oxidation process
usually lasts from several minutes to tens of minutes, whereas the time
scale accessible to MD simulations is typically only picoseconds to
nanoseconds. Several strategies have been proposed to bridge this gap. (1)\
Melt and quench: Crystalline alumina is melted at a high temperature (e.g.,
5000 K), and then cooled down progressively to a low temperature (e.g., 5 K)
to mimic annealing \cite{ref-melt-quench}. This method does produce
amorphous AlO$_{\text{x}}$, but the generated oxide is periodic in all three
dimensions and contains no free surface, making it unsuitable for studying H
atoms bound to the surface. (2) Sequential reactions:\ O$_{\text{2}}$
molecules are introduced to the Al surface one by one to simulate a sequence of
independent chemical reactions \cite{ref-sequential-Cyster}. This method
effectively compresses the waiting time between successive reaction events.
However, because the molecules interact with the surface individually, it
excludes cooperative or co-catalytic processes involving two or more
molecules. In particular, we find that isolated H$_{\text{2}}$O molecules do
not react with the Al surface. (3) High density acceleration: an artificially
high density gas of O$_{\text{2}}$ molecules is placed near the Al surface
so that oxidation proceeds within the nanosecond time scale accessible
to MD \cite{ref-md-grow}. Because the oxidation of the Al surface is
self-limiting, the resulting amorphous AlO$_{\text{x}}$ structure depends
only weakly on the O$_{\text{2}}$ density (see Appendix~\ref{app:NO2}). This approach is adopted in the
MD simulations.

\begin{figure}
	\centering
	\includegraphics[width=8cm]{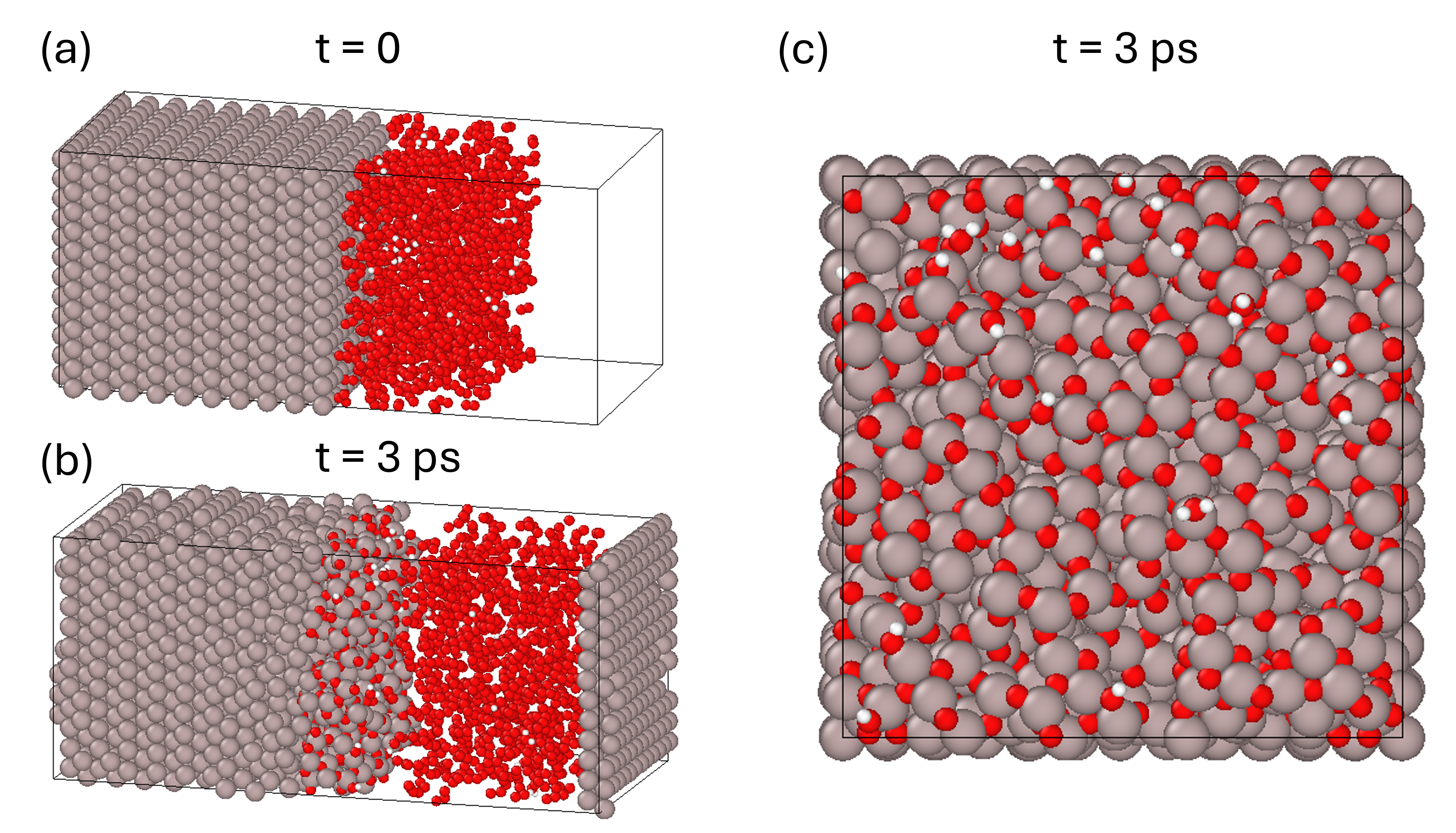}
	\caption{(a) Initial configuration consisting of an Al surface, a gas of O$_{\text{2}}$ and H$_{\text{2}}$O molecules, and a vacuum region. (b) Atomic structure after $3$ ps of MD simulation, showing the formation of an amorphous AlO$_{\text{x}}$ layer on the Al surface. Because of periodic boundary conditions, the initial left-most Al layer appears as the right-most Al layer. (c) Top view of the AlO$_{\text{x}}$ surface at $t=3$ ps. Al, O, and H atoms are represented by gray, red, and white spheres, respectively.}
	\label{fig01}
\end{figure}

The system under investigation consists of an Al surface, a gas of O$_{\text{%
		2}}$ and H$_{\text{2}}$O molecules, and a vacuum region. The Al surface is
modeled as a slab of $20$ fcc(100) layers, each of which contains $12\times 12$
Al atoms. The gas contains $750$ O$_{\text{2}}$ molecules\ and $30$ H$_{%
	\text{2}}$O molecules, initially placed in a region extending from $2$ \AA\ %
to $22$ \AA\ above the top Al surface, leaving a gas-free gap of about $17$~\AA\ between the gas and the periodic image of the bottom Al surface. The numbers of O$_{\text{2}}$ and H$_{\text{2}}$O molecules are chosen so that the resulting AlO$_{%
	\text{x}}$ layer is $\sim 1$ nm thick and contains $\sim 2$ at.\% H atoms. Additional simulations with 10 to 50 H$_2$O molecules show that the number of H atoms incorporated in the oxide increases approximately in proportion to the number of H$_2$O molecules (Appendix~\ref{app:NH2O}).
In total, the system contains $4,470$ atoms, including $2,880$ Al atoms, $%
1,530$ O atoms, and $60$ H atoms. The MD simulations are performed in the
canonical (NVT) ensemble, with fixed number of atoms, volume $34.17\times 34.17\times 78.26$ \AA $^{%
	\text{3}}$, fixed temperature $300$ K, and periodic boundary conditions in all three directions; no atoms are fixed.
Interatomic forces and energies are computed using CHGNet (Crystal Hamiltonian Graph Neural Network)~\cite{ref-chgnet}, a universal machine-learning interatomic potential developed by Deng \emph{et al.} and pretrained on the Materials Project Trajectory dataset, which contains more than 1.5 million atomic configurations with DFT energies, forces, stresses, and magnetic moments. By including magnetic moments as a proxy for atomic charge states, CHGNet captures charge-dependent bonding, which is relevant for the mixed metallic and ionic Al/AlO$_{\text{x}}$ system studied here. The pretrained model (version 0.3.0) is used without further fine-tuning. A machine-learning potential is well suited to the present simulations, which involve the breaking and formation of Al--O, O--H, and Al--H bonds in metallic, ionic, and molecular environments: reactive force fields require system-specific parametrization~\cite{ref-md-grow}, whereas ab initio MD would be prohibitively expensive for 400 simulations of 4{,}470 atoms each. To assess the accuracy of CHGNet for the Al--O system, we benchmarked it against DFT and experiment for crystalline Al and $\alpha$-Al$_2$O$_3$, and against DFT for amorphous Al$_2$O$_3$ bulk and surfaces (Appendix~\ref{app:chgnet}). The lattice parameters agree with those obtained with the Perdew--Burke--Ernzerhof (PBE) functional to within $0.3\%$, and the energies and forces of amorphous Al$_2$O$_3$ are reproduced with mean absolute errors of $16$~meV/atom and $0.19$~eV/\AA, respectively. In addition, the Al--O radial distribution function of the simulated oxide agrees with that of amorphous Al$_2$O$_3$. The MD
simulations run for $3$ ps with a time step of $1$ fs. As a reference, a
typical experimental O$_{\text{2}}$ pressure is $15$ mbar, which corresponds
to $8.46\times 10^{-3}$ O$_{\text{2}}$ molecules in a volume $34.17\times
34.17\times 78.26$ \AA $^{\text{3}}$ at $300$ K. By placing $750$ O$_{\text{2%
}}$ molecules in the same volume, the oxidation process is effectively
accelerated, extending the MD simulation time from $3$ ps to an effective $266$ ns. This scaling follows from kinetic theory: at a fixed temperature, the rate at which gas molecules strike the surface, $\Phi =n\bar{v}/4$ with $n$ the number density and $\bar{v}$ the mean molecular speed, is proportional to $n$, so that requiring the same cumulative exposure $\Phi t$ gives $t_{\text{eff}}=t_{\text{MD}}\,n_{\text{MD}}/n_{\text{exp}}$. The density, rather than the temperature, is increased because a higher temperature would also accelerate thermally activated processes such as diffusion and desorption and thus change the resulting structure. The effective time should therefore be read as an equivalent exposure rather than a physical elapsed time; slow processes such as the relaxation of the oxide are not accelerated. To verify that the resulting oxide does not depend on this choice, we performed additional simulations without H$_2$O, in which the number of O$_2$ molecules is varied from 500 to 800 (Appendix~\ref{app:NO2}). The oxide thickness saturates at about $1$~nm above 650 O$_2$ molecules, so that the 750 O$_2$ molecules used here lie in the saturated regime. Fig. %
\ref{fig01} shows the atomic structures before and after the MD. An
amorphous AlO$_{\text{x}}$ layer forms on the top Al surface, with some H atoms
bound to the oxide surface. The bottom Al surface remains largely unoxidized, because the gas molecules need most of the $3$~ps simulation to cross the gas-free gap. Since no atoms are fixed, the momentum transferred by the gas molecules hitting the top surface moves the slab rigidly by about $2$~\AA\ along $z$ over $3$~ps. 
The resulting oxide thickness of approximately $1$~nm is consistent with typical tunnel barriers in transmon-qubit Josephson junctions, ensuring that the simulated structures are representative of transport-relevant devices. The oxide grows rapidly during the first $\sim 0.3$~ps and saturates at about $1$~nm after $\sim 1.5$~ps, although most of the O$_2$ molecules remain unreacted, reflecting the self-limiting nature of Al oxidation (Appendix~\ref{app:growth}). 


\section{Aluminum oxide structure analysis}

To investigate device-to-device variability, we perform $400$ MD\
simulations starting from different O$_{\text{2}}$ and H$_{\text{2}}$O
molecule configurations. Stoichiometry analysis shows that the resulting
oxide has composition AlO$_{\text{x}}$ with $x=1.25\pm 0.02$, and the
hydrogen concentration is $\left( 2.56\pm 0.54\right) $ at.$\%$, corresponding to about $1.8$ H atoms per nm$^{2}$ of the surface. This is of the same order as the H density of about $0.6$ H atoms per nm$^{2}$ ($0.3$ H$_2$O molecules per nm$^{2}$) estimated for the surface of amorphous alumina exposed to ambient conditions~\cite{ref-Cortie2020}. The
distribution of H atom numbers across $400$ samples is shown in Fig. \ref%
{fig02}. 
If every H atom were incorporated with the same probability $p$, the number of H atoms would follow a binomial distribution. In the simulations, however, the incorporation probability varies from sample to sample, because the oxidation is a self-limiting process in which the reaction rate decreases as the oxide grows in a sample-specific way. Treating $p$ as a random variable that follows a beta distribution, we fit the MD data with a beta-binomial distribution~\cite{ref-betabinomial-skellam,ref-betabinomial-jkk}:
\begin{equation}
	P\left( n\right) =\binom{M}{n}\frac{B\left( n+\alpha ,M-n+\beta \right) }{%
		B\left( \alpha ,\beta \right) },  \label{eq10}
\end{equation}%
where $n$ is the number of H atoms in the oxide of an MD cell with cross section area $A_{1}=34.17\times 34.17$ \AA$^{2}$, $M$ is the number of trials (the maximum value of $n$), $\alpha$ and $\beta$ are shape parameters, and $B\left( x,y\right) $ is the beta function. The fitted parameters are $%
\alpha =17.69$, $\beta =15.36$, and $M=40$. The mean $\langle n\rangle$ and variance $\sigma_{n}^{2}$ of $n$ are
given by
\begin{equation}
	\langle n\rangle =M\frac{\alpha }{\alpha +\beta },
\end{equation}%
and%
\begin{equation}
	\sigma_{n}^{2} =M\left( M+\alpha +\beta \right) \frac{\alpha
		\beta }{\left( \alpha +\beta \right) ^{2}\left( \alpha +\beta +1\right) },
\end{equation}%
which yield $\langle n\rangle =21.41$ and $\sigma_{n}=4.62$. The beta-binomial distribution reduces to the binomial distribution in the limit $\alpha ,\beta \rightarrow \infty $ with $\alpha /(\alpha +\beta )=p$ fixed. For the same mean, a binomial distribution would give $\sigma_{n}=3.15$; the larger value $\sigma_{n}=4.62$ reflects the sample-to-sample variation of the incorporation probability.
%
\begin{figure}
	\centering
	\includegraphics[width=8cm]{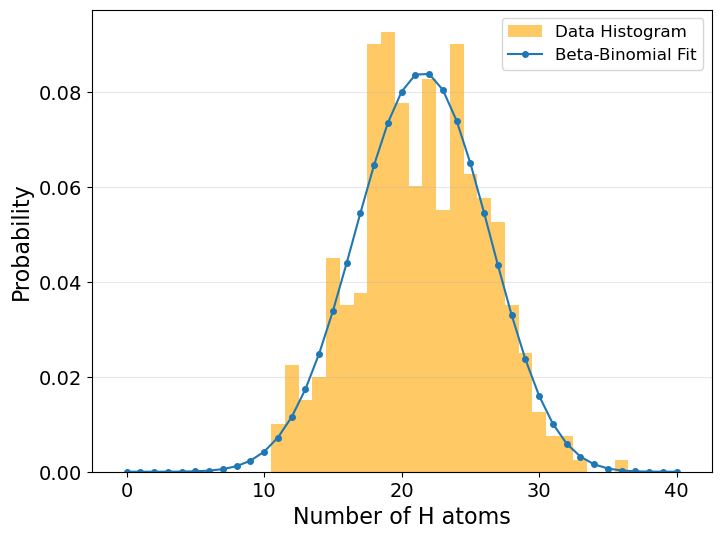}
	\caption{Distribution of H atom numbers in AlO$_{\text{x}}$.}
	\label{fig02}
\end{figure}


Furthermore, we analyzed the binding configurations of H atoms in AlO$_{%
	\text{x}}$. Two atoms are considered bonded if their distance is less than $1.25$ times the sum of their covalent radii~\cite{ref-Cordero2008} ($r_{\text{Al}}=1.21$~\AA, $r_{\text{O}}=0.66$~\AA, $r_{\text{H}}=0.31$~\AA). This gives cutoffs of $1.21$~\AA\ for O--H, $2.34$~\AA\ for Al--O, $1.90$~\AA\ for Al--H, and $1.65$~\AA\ for O--O, which lie between the typical bond lengths ($0.97$~\AA\ for O--H and $1.8$--$2.0$~\AA\ for Al--O in amorphous alumina) and the second-neighbor distances. Hydrogen bonds (H$\cdots$O $\approx 1.6$--$2.0$~\AA) are therefore not counted as bonds. Each H atom is assigned to exactly one motif by the following rules, applied in order of priority: (i) Al-H$_2$O, H on an O atom that carries two H atoms and is bonded to Al; (ii) Al-H-O, H bonded to both O and Al; (iii) Al-OH-Al, H bonded to a single O atom that is bonded to two or three Al atoms (doubly or triply bridging hydroxyl, $\mu_{2}$- or $\mu_{3}$-OH); (iv) Al-OH, H bonded to a single O atom that is bonded to one Al atom (terminal hydroxyl, $\mu_{1}$-OH); (v) Al-O$_2$-H, H bonded to an O--O unit attached to Al; (vi) Al-H-Al and Al-H, H bonded to two or more Al atoms, or to one Al atom, and to no O atom; (vii) interstitial, H with no bonded neighbors; (viii) Al-O-H-O-Al, H bonded to two O atoms that are each bonded to Al. The hydroxyl motifs follow the standard classification of surface hydroxyl groups on alumina by the number of Al atoms bonded to the O atom~\cite{ref-Knozinger1978,ref-Digne2004}. Surface sites are identified by lowering a probe sphere of radius $4$~\AA\ onto the oxide on a $100\times 100$ lateral grid; an H atom is classified as surface-bound if its bonded O or Al atom is touched by the probe. About $91.3\%$ of the H atoms form hydroxyl (-OH) groups bonded to Al: terminal $\mu_{1}$-OH groups (Al-OH, $37.0\%$) and bridging hydroxyl groups (Al-OH-Al, $54.3\%$). The bridging groups are mostly doubly bridging $\mu_{2}$-OH, while triply bridging $\mu_{3}$-OH groups account for only $0.7\%$ of the H atoms. All three hydroxyl types are known on alumina surfaces~\cite{ref-Knozinger1978,ref-Digne2004}, where the number of Al atoms bonded to the hydroxyl O atom determines the OH stretching frequency and the acid--base properties of the group~\cite{ref-Digne2004}. Another $%
5.4\%$ of the H atoms form Al-H$_{\text{2}}$O motifs, where H atoms bind to
Al via H$_{\text{2}}$O molecules. Other motifs are rare, including Al-O$_{%
	\text{2}}$-H, Al-H, Al-H-Al, Al-H-O, interstitial sites, and Al-O-H-O-Al.
Notably, most H atoms are located near the surface of the AlO$_{\text{x}}$
layer. Because the local bonding determines the potential-energy landscape of an H atom, these motifs provide a natural set of units for identifying candidate H-related TLS. For example, the H atom of a terminal Al-OH group may rotate about the Al--O axis, and in the rare Al-O-H-O-Al motif an H atom is shared almost symmetrically between two O atoms (O--H distances of about $1.1$--$1.2$~\AA), which is the O--H$\cdots$O configuration proposed as a hydrogen-tunneling TLS~\cite{Gordon2014TLS}.
The detailed statistics of the binding motifs are summarized below.
\newcommand{\motif}[1]{\raisebox{-0.5\height}{\includegraphics[height=1.15cm]{figures/#1}}}
\begin{table*}[t]
	\centering
	\caption{Statistics (mean $\pm$ standard deviation) of hydrogen-related motifs in AlO$_{\text{x}}$ and the probabilities of binding to surface sites. Representative atomic structures of the motifs are taken from the final MD configurations. Al, O, and H atoms are represented by gray, red, and white spheres, respectively.}
	\label{tab01}
	\setlength{\tabcolsep}{2.5pt}
	\renewcommand{\arraystretch}{1.3}
	\footnotesize
	\begin{tabular}{|l|c|c|c|c|c|c|c|c|c|}
		\hline
		\textbf{Motif} & Al-OH & Al-OH-Al & Al-H$_2$O & Al-O$_2$-H & Al-H & Al-H-Al & Al-H-O & interstitial & Al-O-H-O-Al \\ \hline
		\textbf{\shortstack[l]{Atomic\\structure}} & \motif{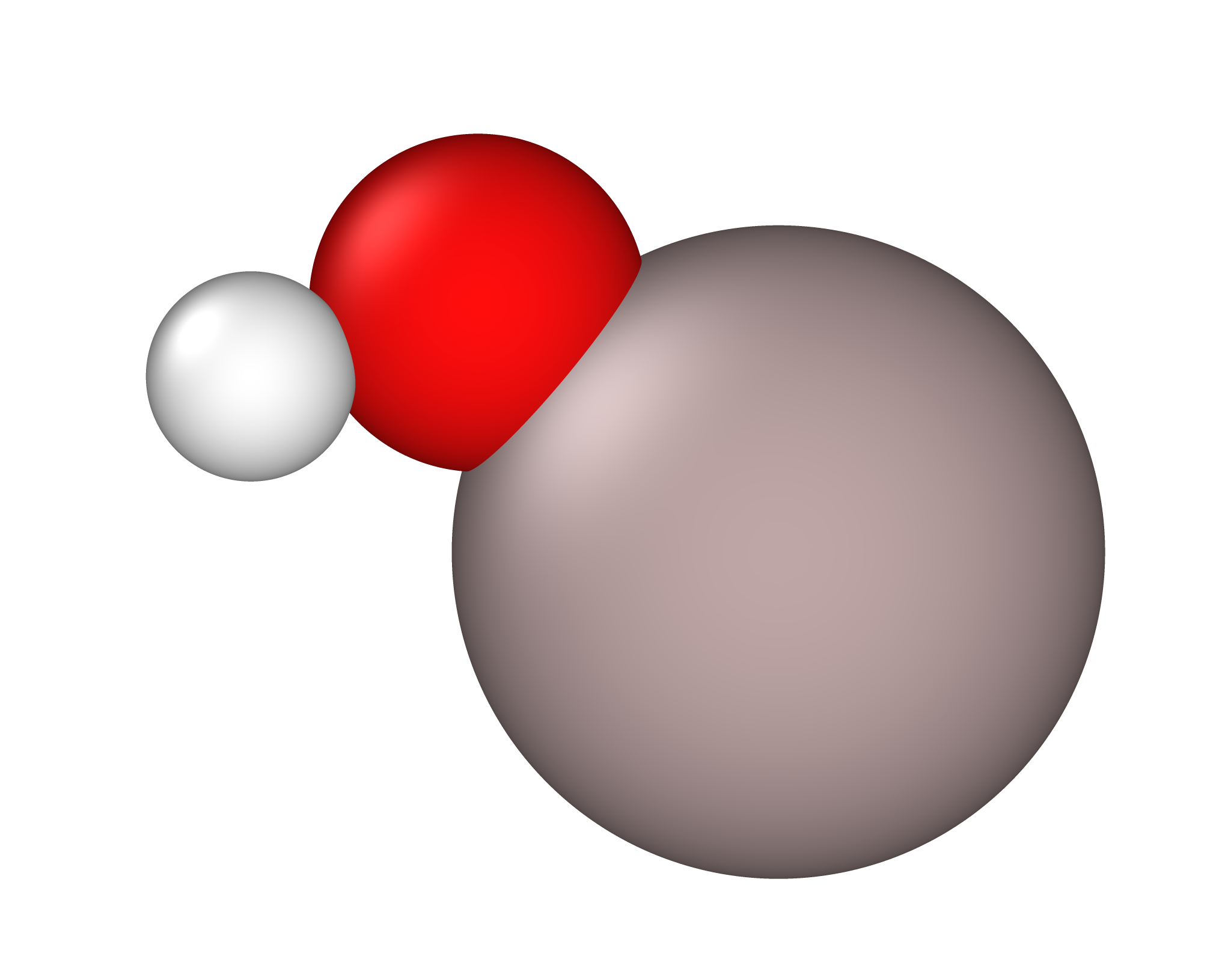} & \motif{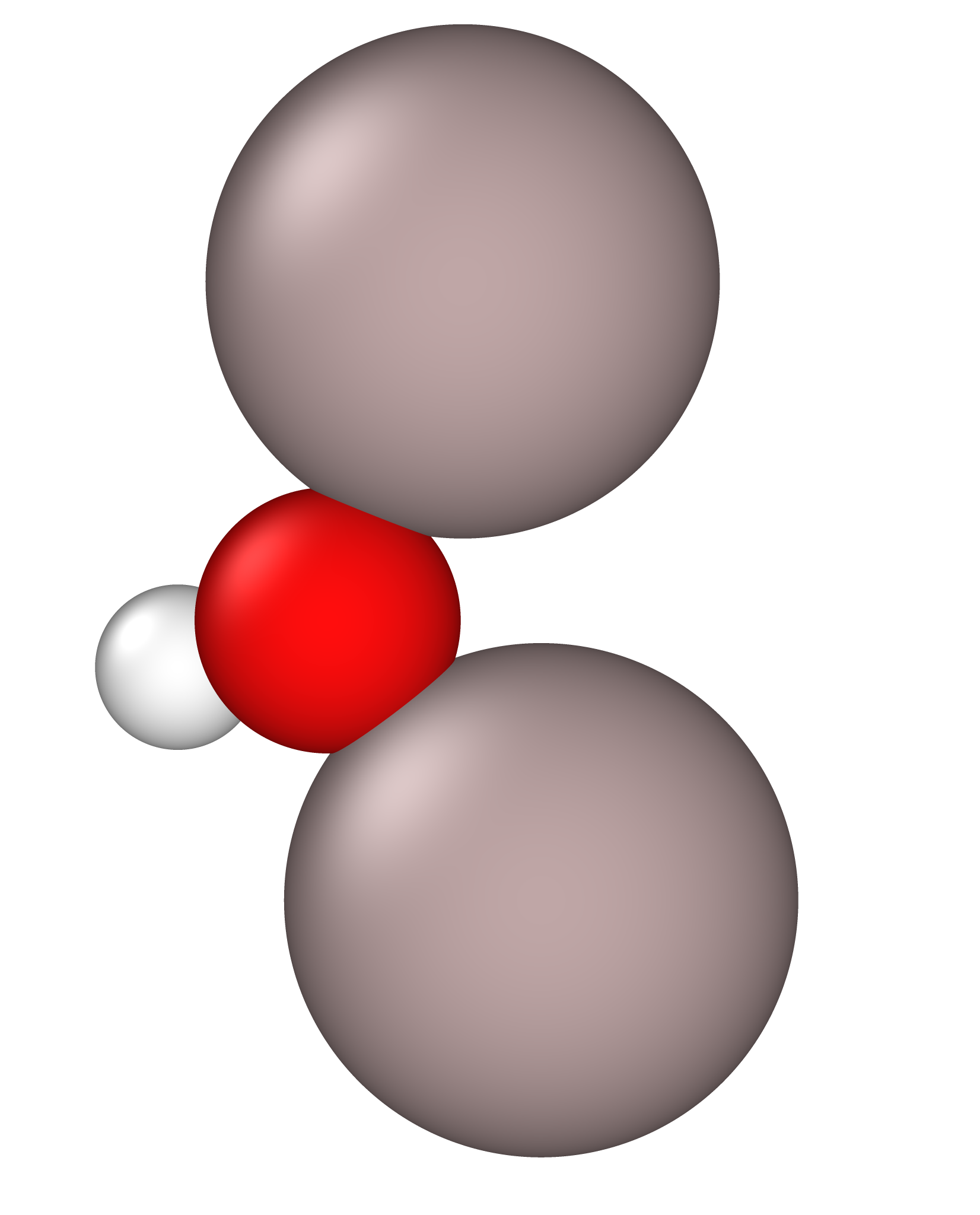} & \motif{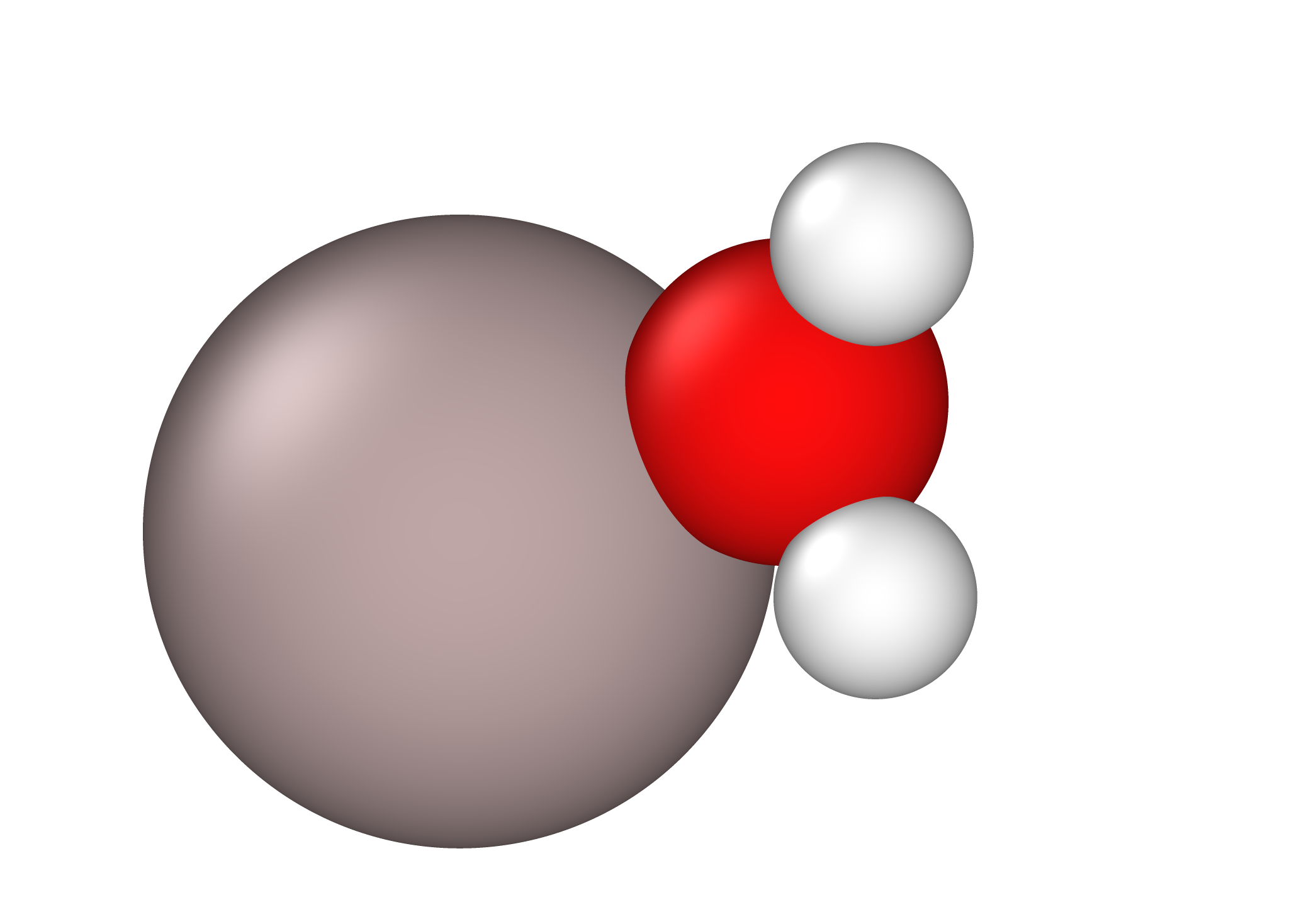} & \motif{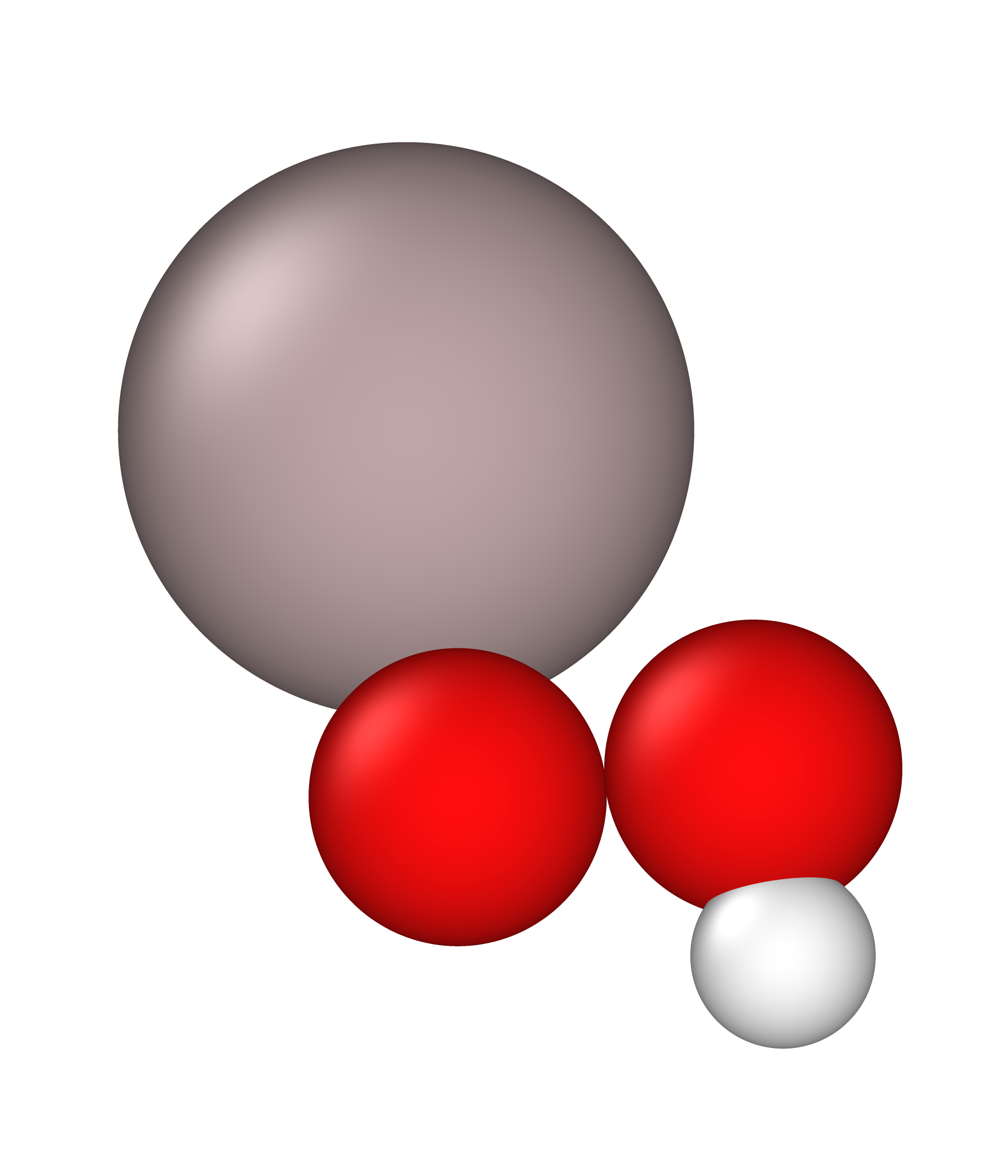} & \motif{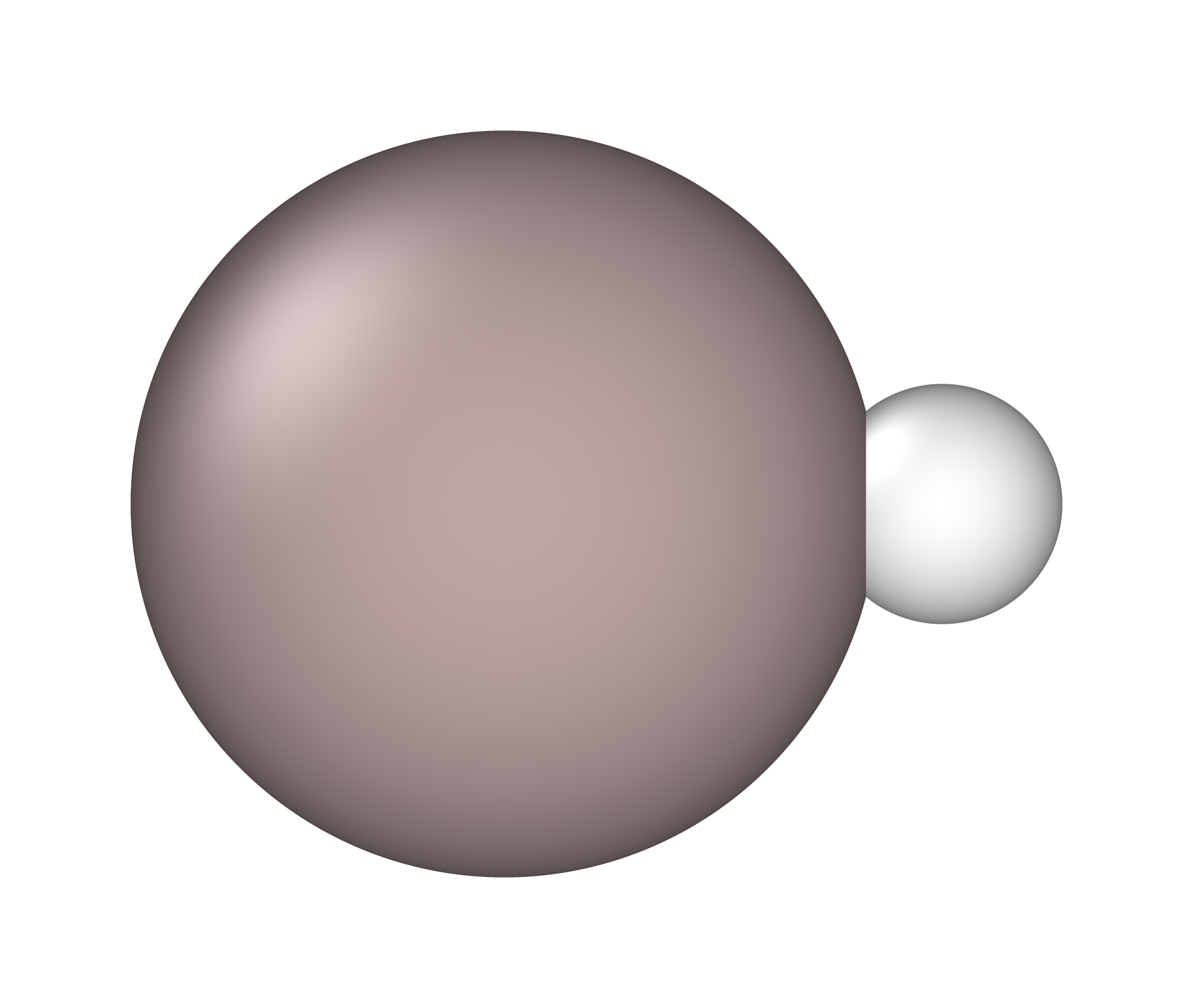} & \motif{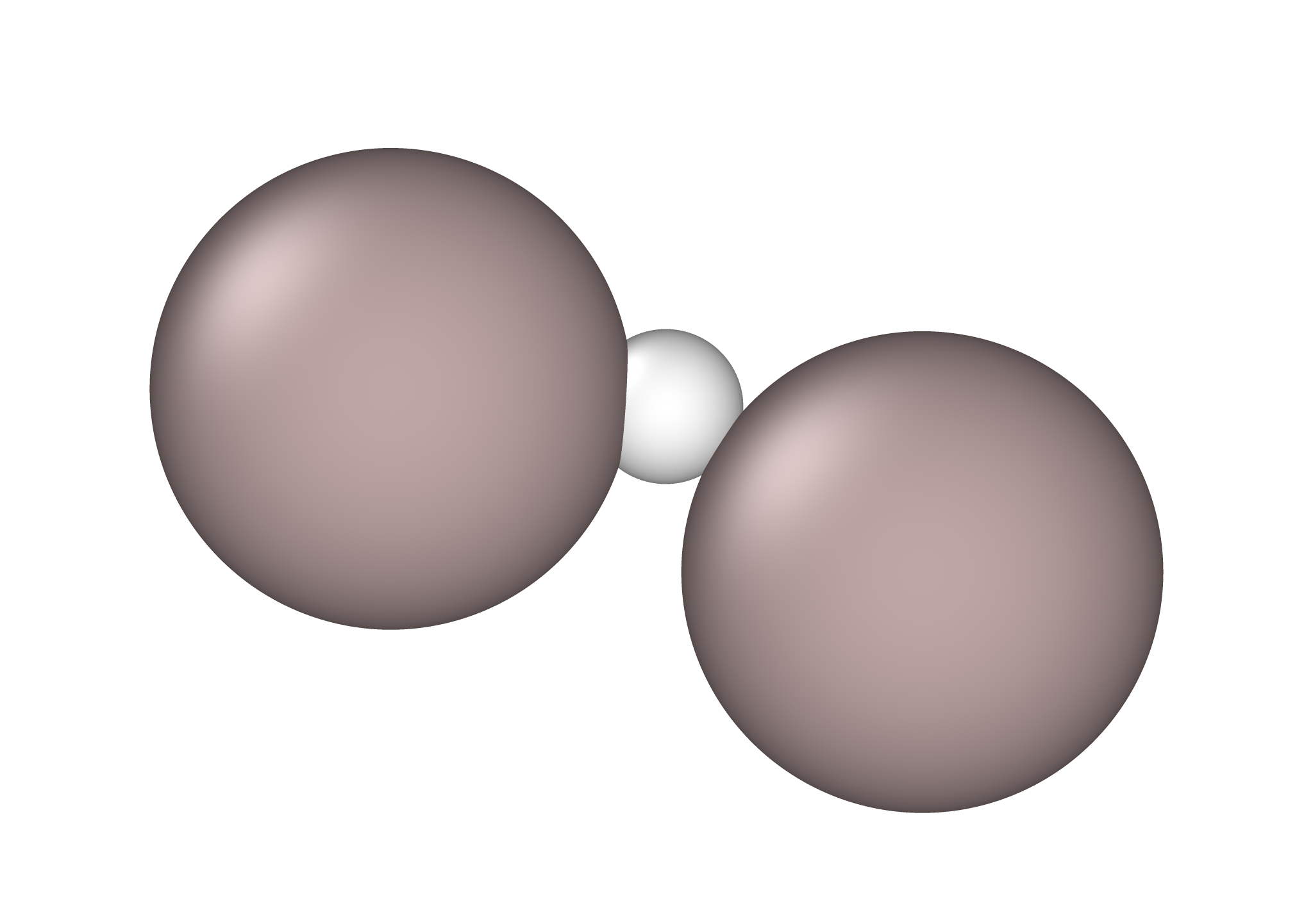} & \motif{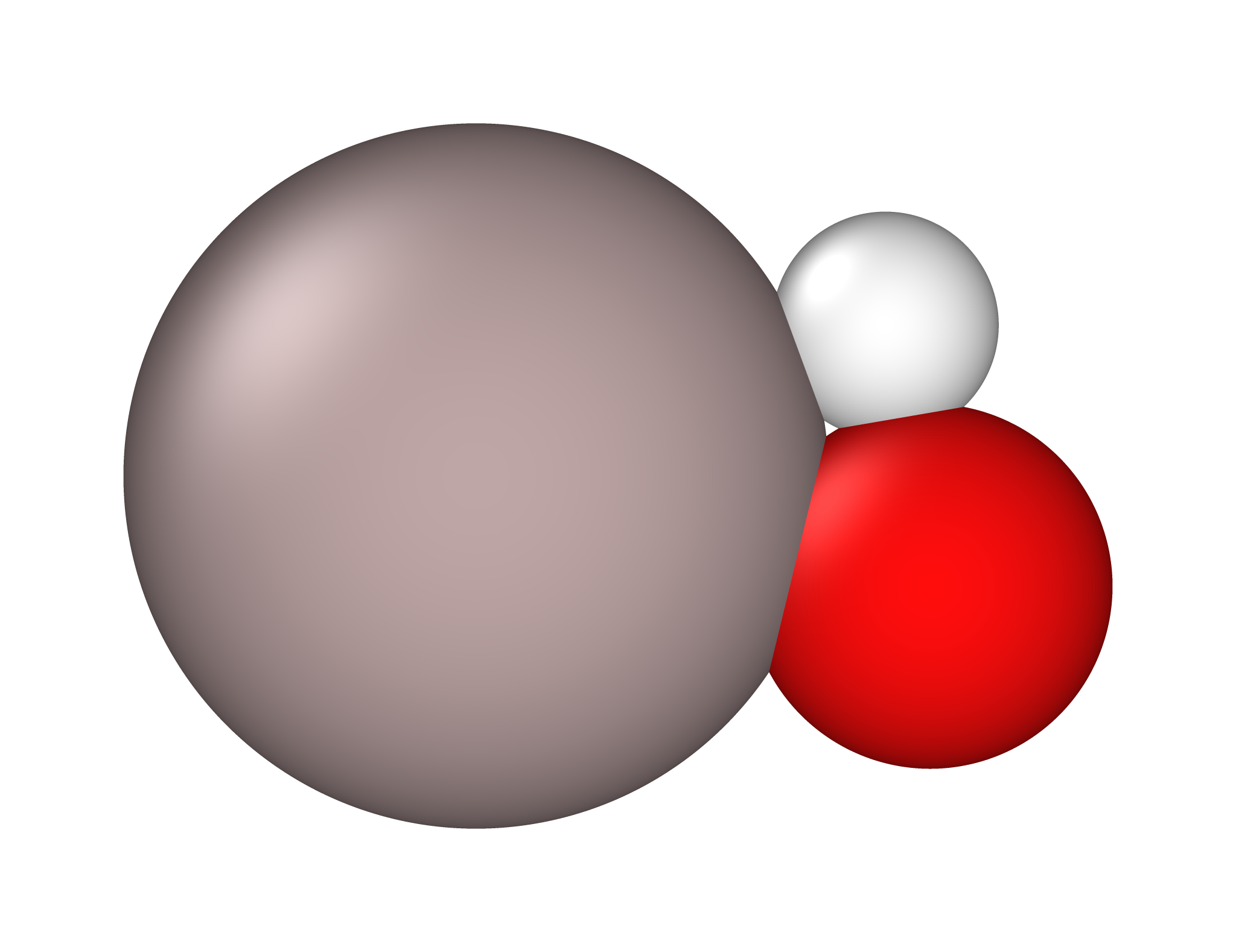} & \motif{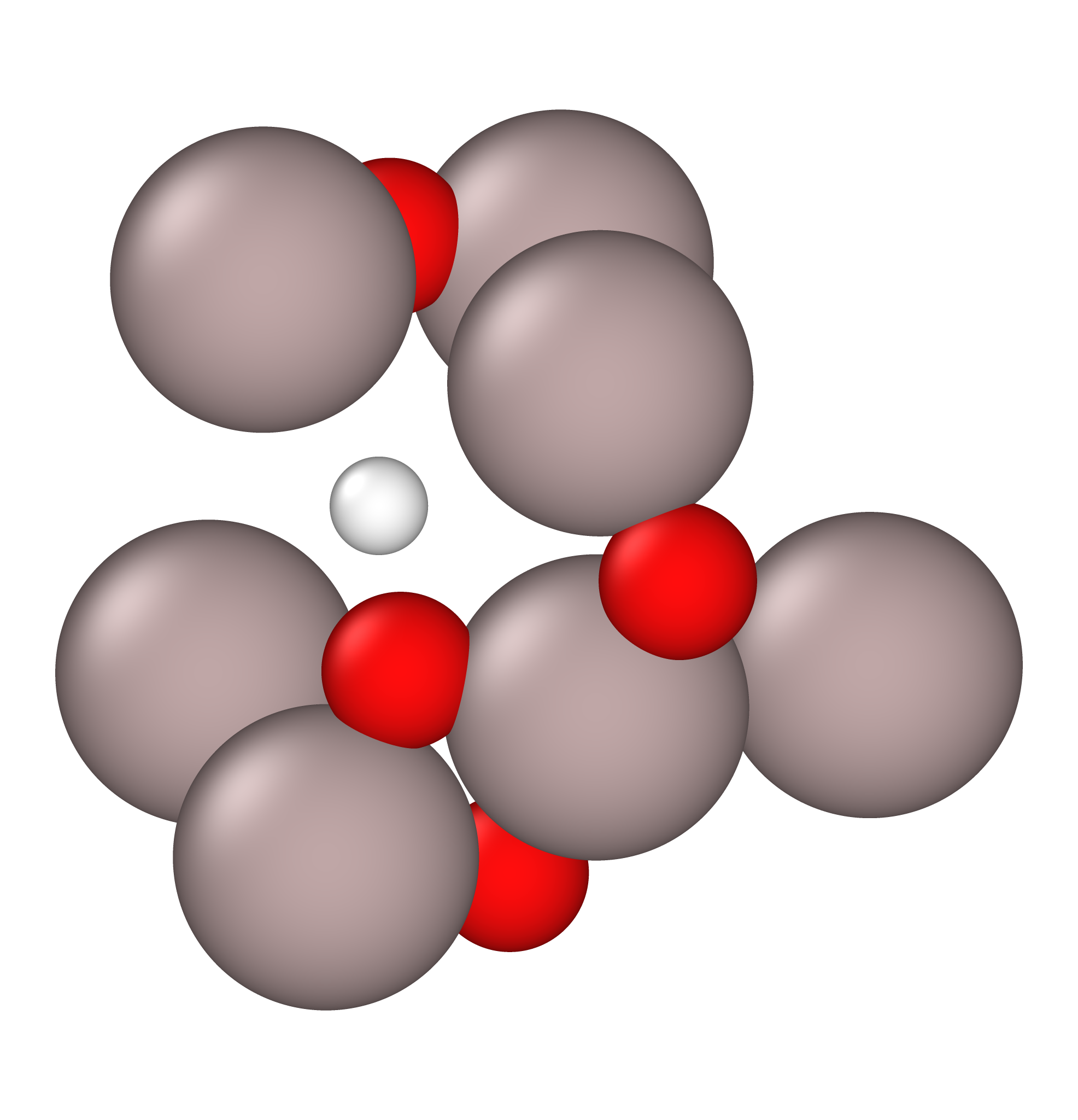} & \motif{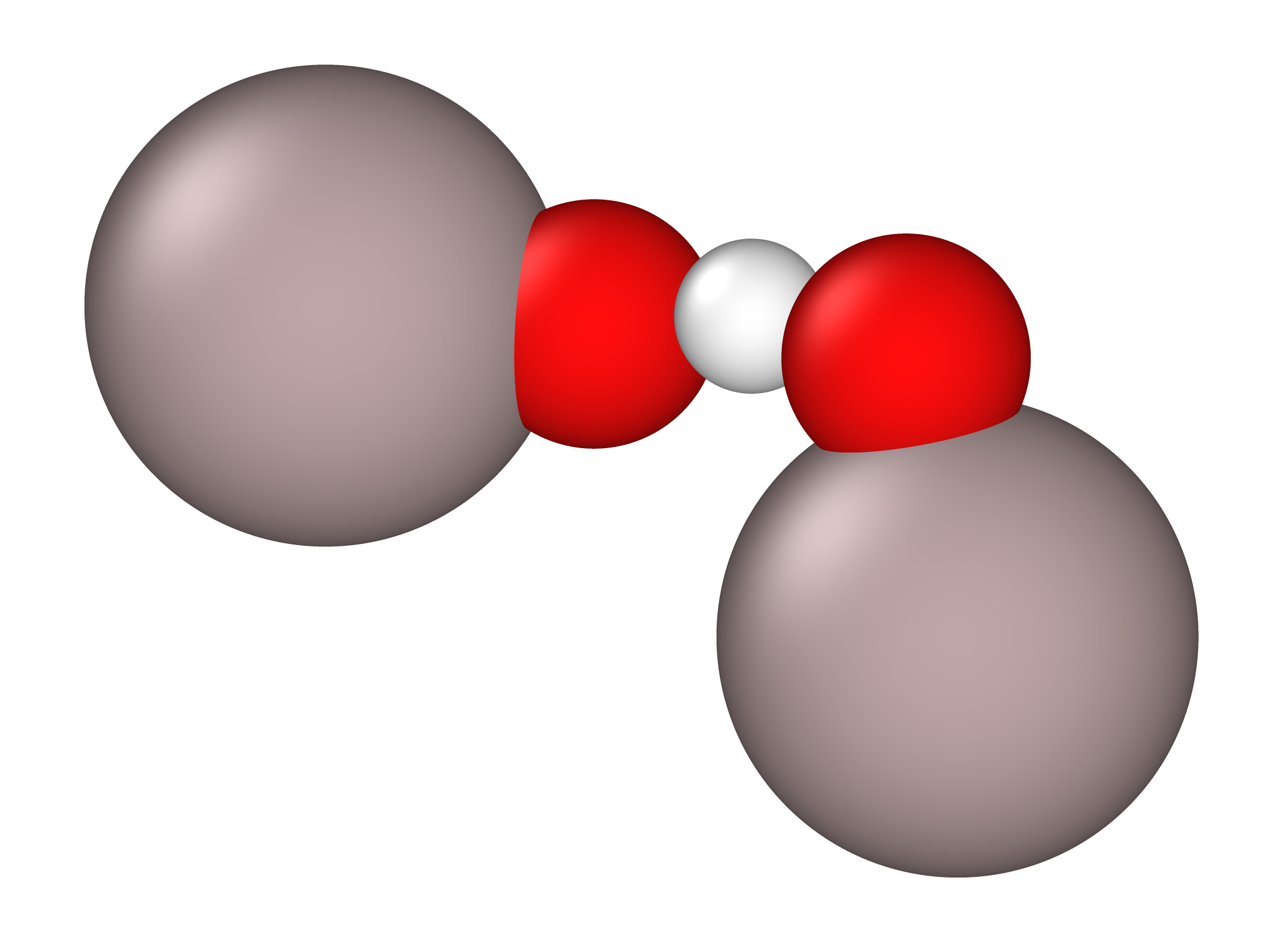}\,\motif{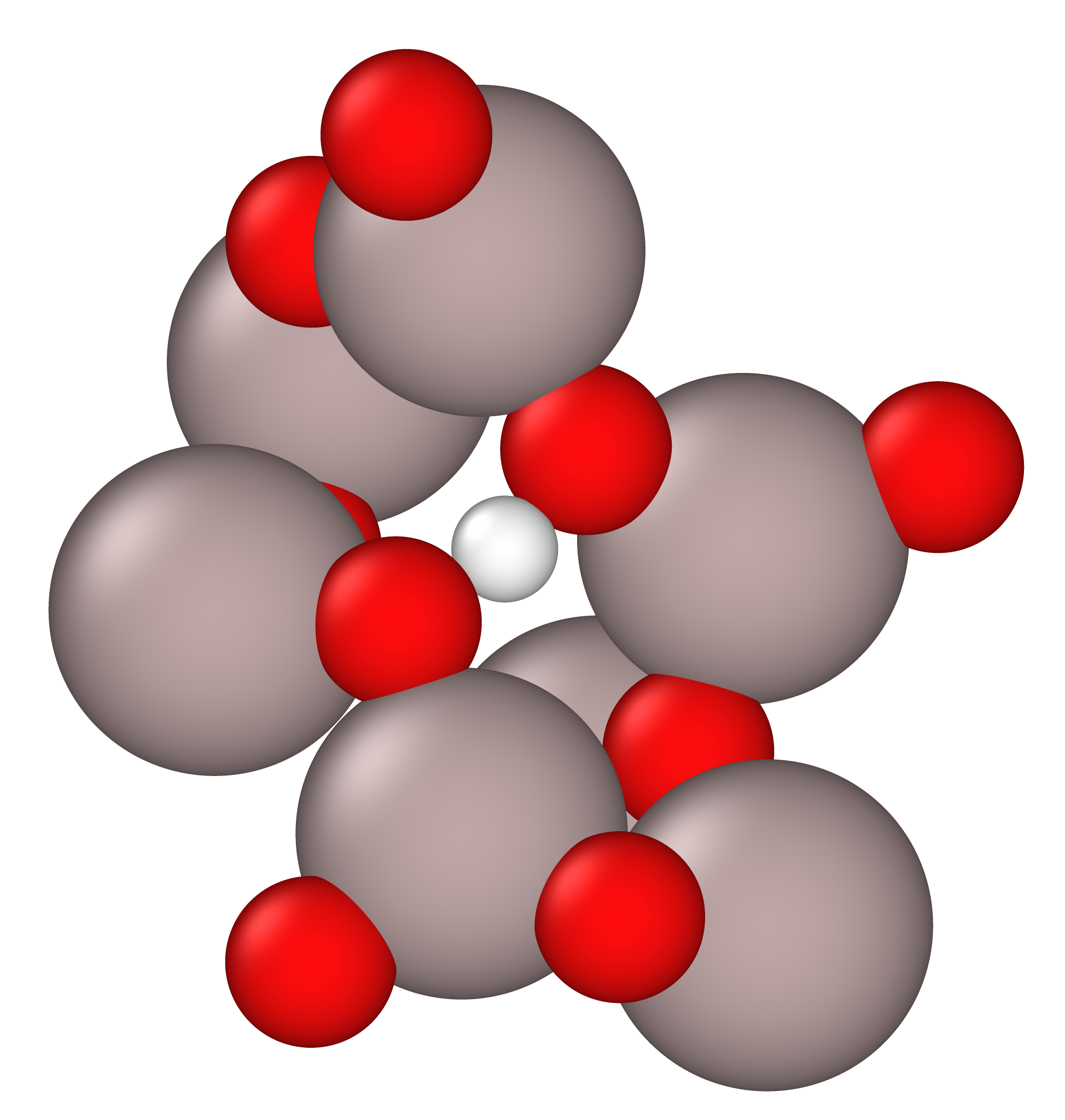} \rule[-0.75cm]{0pt}{1.5cm} \\ \hline
		\textbf{Binding type (\%)} & $37.0 \pm 10.8$ & $54.3 \pm 10.8$ & $5.4 \pm 6.8$ & $0.7 \pm 1.8$ & $1.3 \pm 2.6$ & $0.4 \pm 1.3$ & $0.6 \pm 1.5$ & $0.4 \pm 1.3$ & $0.0 \pm 0.3$ \\ \hline
		\textbf{Surface site (\%)} & $94.8$ & $76.5$ & $44.6$ & $13.8$ & $2.5$ & $0.0$ & $9.0$ & $0.0$ & $0.0$ \\ \hline
	\end{tabular}
\end{table*}

\section{Quantum transport calculation}

Next, we perform quantum transport calculations for Josephson junctions to
investigate the impact of hydrogen contamination on the Josephson energy.
Approaches to computing transport through Al/AlO$_{\text{x}}$/Al junctions fall into three broad classes. (i) Phenomenological models: the barrier is treated as a one-dimensional potential, the tunneling current is obtained from the Wentzel--Kramers--Brillouin (WKB) approximation~\cite{ref-Simmons1963}, and $I_{c}$ then follows from the Ambegaokar--Baratoff relation~\cite{ambegaokar1963tunneling}. Such models are efficient but cannot capture the three-dimensional amorphous structure of the oxide. (ii) Atomistic models: Cole and co-workers constructed three-dimensional atomistic models of amorphous Al--AlO$_{\text{x}}$--Al barriers~\cite{ref-DuBois2016,ref-sequential-Cyster} and simulated their transport with the NEGF formalism~\cite{ref-Cyster2020}. They showed that variations in the local potential landscape create channels that dominate conduction, and that oxygen-deficient barriers can contain metallic channels~\cite{ref-Cyster2020}. In later work, they showed that pinholes, which form when the oxidation is incomplete, and weak points in continuous barriers give rise to excess and leakage currents and to deviations from the sinusoidal current--phase relation~\cite{ref-Bayros2024}. Kim, Ray, and Lordi generated junction models with ab initio molecular dynamics and computed Cooper-pair tunneling rates from the three-dimensional electrostatic potential, revealing a channeling effect on $I_{c}$~\cite{ref-Kim2020}. (iii) First-principles NEGF-DFT, which treats the electronic structure and the open-boundary transport self-consistently: it has been applied to the effects of the interface termination~\cite{ref-Shan2022}, the oxide stoichiometry~\cite{ref-JJ-stoichiometry-nanodcal}, and interface disorder~\cite{ref-Han2026} on junction transport, as well as to the difference between amorphous and crystalline barriers~\cite{ref-Zhang2026}. The present work belongs to class (iii) and applies NEGF-DFT to the effect of hydrogen impurities.

Since the contamination effect depends on details at the atomic scale, it is
necessary to apply density functional theory (DFT) to solve the Hamiltonian
of the system. Moreover, a Josephson junction is a two-probe system, in
which a central scattering region is connected to two semi-infinite leads.
The nonequilibrium Green's function (NEGF) formalism provides a powerful
theoretical framework for describing such open systems. The NanoDCAL software
package implements the NEGF-DFT theory \cite{ref-NEGF-DFT} and enables
atomistic modeling of nanoscale devices from first principles \cite%
{ref-nanodcal}. In this work,
we use NanoDCAL to compute the tunneling current through Josephson junctions
with or without H atoms.

To begin, we construct atomistic models of Al/AlO$_{\text{x}}$/Al Josephson junctions. 
The amorphous AlO$_{\text{x}}$ barrier is approximated by crystalline $\alpha$-Al$_2$O$_3$. This choice is motivated by three considerations. First, the barrier structure strongly affects the absolute transport properties: amorphous Al$_2$O$_3$ barriers give a higher and more variable $E_{J}$ than crystalline barriers of the same thickness, because compositional inhomogeneity creates preferential tunneling paths~\cite{ref-Zhang2026,ref-Cyster2020}. The present work, however, concerns the relative change in the transmission caused by an H atom at the interface, which we expect to be less sensitive to the long-range order of the oxide than the absolute transmission; we adopt this as an assumption of the model. Second, in amorphous barriers the transmission varies strongly from one configuration to another~\cite{ref-Zhang2026}, much more than the change of a few percent caused by a single H atom. Resolving the H effect would therefore require paired calculations, with and without H, on the same amorphous configuration, repeated over many configurations and H sites, whereas a crystalline barrier provides a well-defined reference in which the H effect can be determined unambiguously. Third, an amorphous junction of realistic thickness contains more than $10^{3}$ atoms~\cite{ref-Zhang2026}; self-consistent NEGF-DFT calculations at this size are feasible for a few configurations, but not for the many configurations and H sites that would be required. Consequently, the absolute values of $T(E_{F})$ and $E_{J}$ obtained here refer to an idealized crystalline barrier, whereas the quantity of interest is the relative change caused by hydrogen. Calculations with amorphous barriers and different H sites are a natural extension of this work.
The central scattering region consists of $12$ fcc(111) Al layers on the left, $12$ alternating Al and O layers of $\alpha $-Al$_{\text{2}}$O$_{\text{3}}$ in the middle, and $13$ fcc(111) Al
layers on the right. Here, the Al layers are compressed in the transverse directions to $96.7\%$ of the equilibrium lattice constant, so that they match the lattice of $\alpha $-Al$_{\text{2}}$O$_{\text{3}}$, resulting in a common cross section $9.61$ \AA $\ \times 8.32$ 
\AA . According to the MD simulations, H atoms are most likely to bind to
surface O atoms. To simulate the contamination effect, a single H atom is
attached to an O atom on the surface of $\alpha $-Al$_{\text{2}}$O$_{\text{3}%
}$ (120 atoms), which corresponds to a hydrogen concentration of $%
1/120\approx 0.83$ at.$\%$. As a result, the central scattering region
contains $420$ or $421$ atoms without or with the H atom. The corresponding
two-probe systems are referred to as JJ and JJ-H, respectively. The system is
periodic in the $x$ and $y$ directions, and transport is along the $z$
direction.

\begin{figure}
	\centering
	\includegraphics[width=9cm]{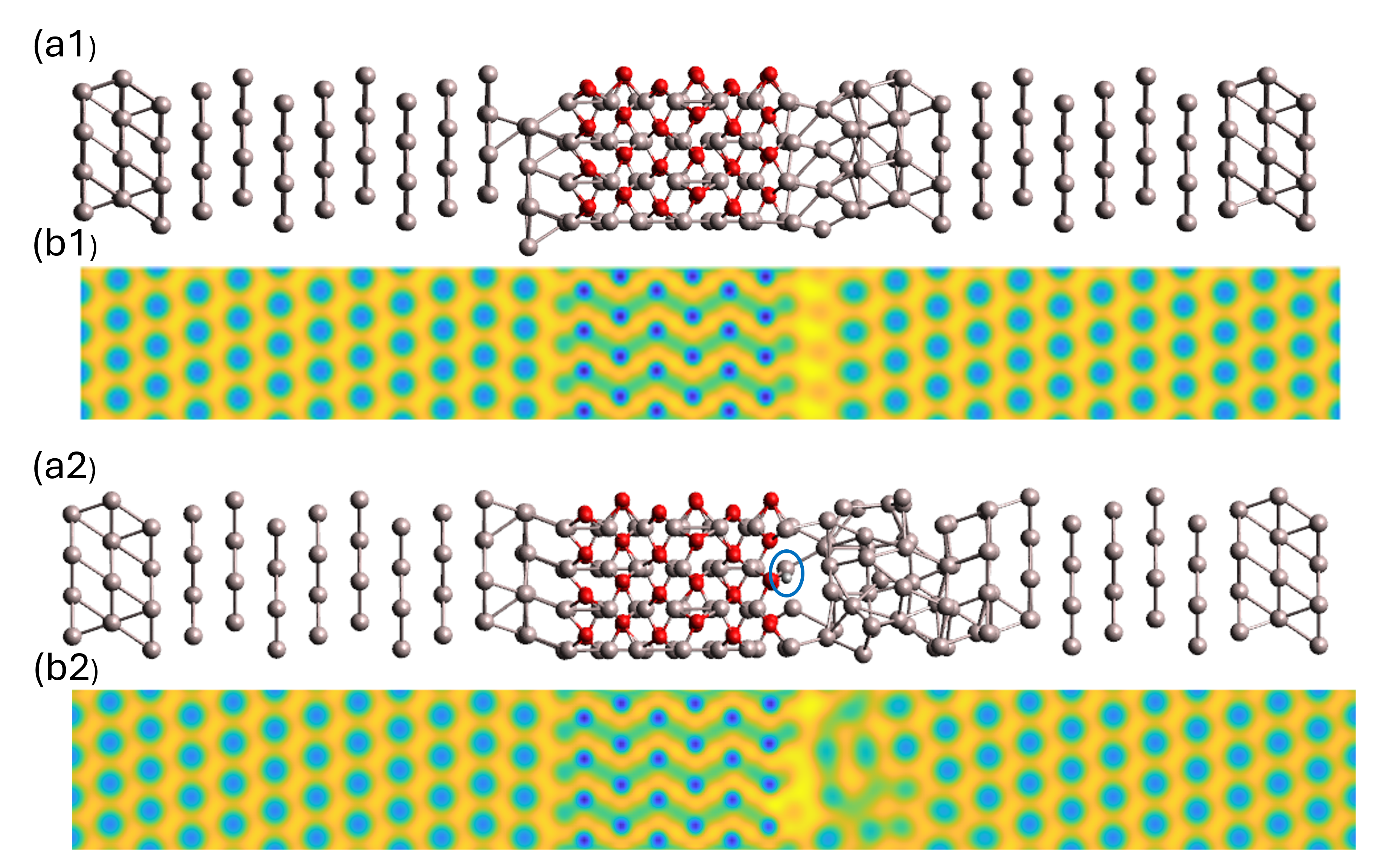}
	\caption{(a1) and (a2): Relaxed atomic structures of JJ and JJ-H. The H atom is highlighted by a blue circle. (b1) and (b2): Corresponding averaged potentials in the $xz$ plane obtained from NanoDCAL self-consistent calculations. The yellow region indicates an area of low potential, while the blue region indicates an area of high potential.
}
	\label{fig03}
\end{figure}

The two-probe structures are optimized using CHGNet through a two-step
relaxation procedure. Before the relaxation, the distances between the Al
layers and $\alpha $-Al$_{\text{2}}$O$_{\text{3}}$ layers are set to $1.9$ 
\AA , providing a reasonable initial guess. In the first relaxation step,
the three leftmost Al layers are fixed while the rest of the system is
allowed to relax. After this step, the three rightmost Al layers are
replaced by lead layers. In the second relaxation step, both the three
leftmost and three rightmost Al layers are fixed, and the remaining atoms
are relaxed. After the relaxation, the three leftmost and three rightmost Al
layers are identical to lead layers, while the distances between Al layers
and $\alpha $-Al$_{\text{2}}$O$_{\text{3}}$ layers are determined naturally
by energy minimization. The resulting structures are shown in Fig. \ref%
{fig03}a1 and \ref{fig03}a2 for JJ and JJ-H, respectively. It is evident
that introducing a single H atom at the Al/Al$_{\text{2}}$O$_{\text{3}}$
interface significantly alters the local atomic structure.


Given the atomic structure, the Hamiltonian of the Josephson junction is
solved self-consistently using NanoDCAL. In the calculations, double-zeta
plus polarization (DZP) atomic orbitals are used as the basis set, where
each Al atom is described by $13$ orbitals (s1, s2, p1, p2, d), each O atom
by $13$ orbitals (s1, s2, p1, p2, d), and each H atom by $5$ orbitals (s1,
s2, p). The basis set is optimized to ensure that the band structure
obtained from atomic orbitals closely reproduces that obtained from plane
waves. The real-space grid\ resolution is determined by the energy cutoff $%
1000$ eV, and the $k$-sampling in reciprocal space is set to $4\times 5$ in
the $x$ and $y$ directions for the self-consistent calculations. The transmission coefficients at the Fermi energy are evaluated with a denser $k$-sampling of $40\times 50$.

The generalized gradient approximation (GGA) is adopted as the exchange-correlation (XC) functional. However, GGA is known to underestimate the band gaps of insulators, which in turn leads to inaccurate band alignment. Internal photoemission measurements locate the Al Fermi level $2.9\pm0.2$~eV below the conduction-band minimum of Al$_2$O$_3$~\cite{ref-band-offset-exp}, and DFT calculations give $2.80$~eV before band-gap correction~\cite{ref-band-offset-thy}. To remedy this problem, we employ GGA+$U$, with an on-site $U$ correction applied to the p orbitals of O atoms. We find that $U=1.0$ eV yields the correct band alignment between Al and Al$_2$O$_3$; the dependence of the band gap and the transmission coefficients on $U$ is discussed in Appendix~\ref{app:U}. 
Accurate band alignment is essential for transport calculations because the tunneling current depends exponentially on the barrier height seen by electrons at the Fermi level; even modest errors in the Fermi level position relative to the band edges can lead to orders-of-magnitude changes in the transmission coefficient. 

\begin{figure}
	\centering
	\includegraphics[width=\columnwidth]{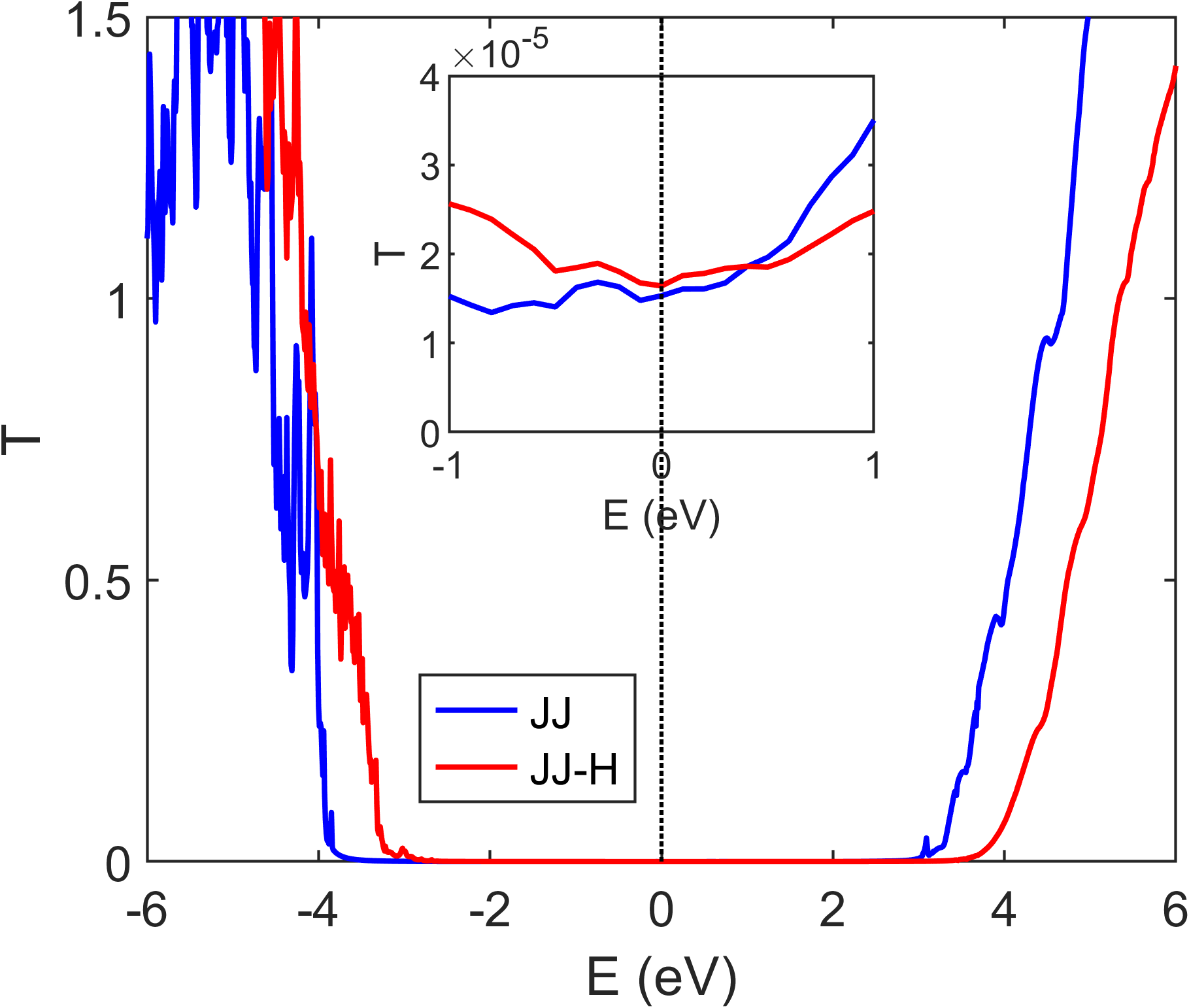}
	\caption{Transmission coefficients for JJ\ (blue curve) and JJ-H (red curve), respectively. The Fermi energy is indicated by the vertical dotted line. The inset shows a magnified view of $T(E)$ in the energy window $-1~\text{eV}<E<1~\text{eV}$ around the Fermi energy, computed with a denser $k$-sampling of $40\times 50$.}
	\label{fig04}
\end{figure}

Fig. \ref{fig03}b1 and \ref{fig03}b2 show the self-consistent potentials JJ
and JJ-H, respectively. The potential includes the Hartree potential, the XC
potential, and the local part of the pseudopotential, averaged along the $y$
direction. The yellow areas indicate regions with lower potential, where
electrons have a higher probability to tunnel through. Fig. \ref{fig04}
shows the transmission coefficients $T\left( E\right) $ for JJ and JJ-H. As
expected, $T\left( E\right) $ is exponentially small inside the band gap. At
the Fermi energy $E_{F}$, the transmission coefficients are $T_{0}^{\text{JJ}}\left(
E_{F}\right) =1.531\times 10^{-5}$ and $T_{0}^{\text{JJ-H}}\left( E_{F}\right) =1.641\times 10^{-5}$ for JJ and JJ-H, respectively, for the cross section area $A_{0}=$ $%
9.61\times 8.32$ \AA $^{\text{2}}$. Hydrogen contamination shifts $T\left(
E\right) $ by about $0.8$ eV such that $E_{F}$ is closer to the valence band
edge, thereby slightly increasing the transmission coefficient at $E_F$ by $7.2\%$. Mulliken analysis shows that the H atom carries a net charge of $+0.56e$ and that the O atom bonded to it gains about $0.1$ electrons relative to the other O atoms in the same layer. The resulting interface dipole, with its positive end toward the Al electrode, raises the bands of Al$_2$O$_3$ relative to the Fermi level of the electrodes.


\section{Device-to-device variability}

The Josephson energy $E_{J}$ is a key parameter of Josephson junctions in
superconducting qubit applications. $E_{J}$ is related to the critical
current $I_{c}$ by%
\begin{equation}
	E_{J}=\frac{\hbar }{2e}I_{c}.  \label{eq11}
\end{equation}%
In the tunneling limit, $I_{c}$ can be calculated using the
Ambegaokar-Baratoff relation%
\begin{equation}
	I_{c}=\frac{\pi \Delta }{2eR_{N}},  \label{eq12}
\end{equation}%
where $R_{N}$ is the normal-state resistance and $\Delta $ is the
superconducting gap. $R_{N}$ is related to the conductance $G$ by%
\begin{equation}
	R_{N}=\frac{1}{G},  \label{eq13}
\end{equation}%
where 
\begin{equation}
	G=\frac{2e^{2}}{h}T\left( E_{F}\right) .  \label{eq14}
\end{equation}%
Combining Eqs. (\ref{eq11},\ref{eq12},\ref{eq13},\ref{eq14}) yields%
\begin{equation}
	E_{J}=\frac{\Delta }{4}T\left( E_{F}\right) =\frac{\Delta }{4}\frac{%
		T_{0}\left( E_{F}\right) }{A_{0}}A,  \label{eq15}
\end{equation}%
where $A$ is the cross section area of the Josephson junction. Here, $%
T_{0}\left( E_{F}\right) $ and $A_{0}$ denote the transmission coefficient
and the cross section area in the transport calculation.

The two parts of this work are connected through the Josephson energy of a realistic junction. The MD simulations (Secs.~II and III) provide the number of H atoms per unit area, its statistical distribution, and the preferred binding sites of the H atoms, which are mainly O atoms at the oxide surface. The transport calculations (Sec.~IV) provide the change in the tunneling transmission caused by an H atom at such a site. Neither part alone determines $E_{J}$; together they give its mean and its device-to-device spread. To calculate $E_{J}$ of a Josephson junction containing $N$ H atoms, we assume that each H atom changes the transmission by the same amount as in the JJ-H model, so that the contributions of the H atoms are additive. This assumption is reasonable for the present system, because the H atoms in the JJ-H model form a periodic array with a mean spacing of $8.9$~\AA\ (one H atom per $80$~\AA$^{2}$), comparable to the mean H--H spacing of about $7.4$~\AA\ in the simulated oxide, and because the change in the transmission coefficient caused by an H atom is small ($7.2\%$), so that the contributions of individual H atoms add linearly to first order.
Under this assumption, $E_{J}$ is obtained as%
\begin{equation}
	E_{J}=\left( A-NA_{0}\right) \frac{E_{J,\text{JJ}}}{A}+NA_{0}\frac{E_{J,%
			\text{JJ-H}}}{A},  \label{eq16}
\end{equation}%
where $N$ is the number of H atoms in the junction, and $E_{J,\text{JJ}}$ and $E_{J,\text{JJ-H}}$ are the Josephson energies of a junction of area $A$ obtained from Eq.~(\ref{eq15}) with the transmission coefficients $T_{0}(E_{F})$ of JJ and JJ-H, respectively. For $A=200\times 200$ nm$^{2}$, $E_{J,\text{JJ}}/h=9.26$ GHz and $E_{J,\text{JJ-H}}/h=9.92$ GHz. Fluctuations in $N$ lead
to the device-to-device variability of $E_{J}$. 
Equation~(\ref{eq16}) can be rewritten as
\begin{equation}
	E_{J}=E_{J,\text{JJ}}+N\,\Delta E,
	\label{eqEJadd}
\end{equation}
where
\begin{equation}
	\Delta E\equiv \frac{A_{0}}{A}\left( E_{J,\text{JJ-H}}-E_{J,\text{JJ}}\right) .
	\label{eqDe}
\end{equation}
Each H atom thus changes $E_{J}$ by a fixed amount $\Delta E$, independent of the junction area. Since the junction consists of $K=A/A_{1}$ statistically independent patches, each the size of the MD cell of area $A_{1}$, $N=\sum_{i=1}^{K}n_{i}$, with each $n_{i}$ distributed according to Eq.~(\ref{eq10}). Hence $\langle N\rangle =K\langle n\rangle$ and $\sigma_{N}=\sqrt{K}\,\sigma_{n}$. For $K\gg 1$, the central limit theorem gives a Gaussian distribution
\begin{equation}
	\tilde{P}(E_{J})=\frac{1}{\sqrt{2\pi }\,\sigma_{E_{J}}}
	\exp \!\left[ -\frac{\left( E_{J}-\langle E_{J}\rangle \right)^{2}}{2\sigma_{E_{J}}^{2}}\right] ,
	\label{eqPEJ}
\end{equation}
with
\begin{equation}
	\langle E_{J}\rangle =E_{J,\text{JJ}}+K\langle n\rangle \,\Delta E,\qquad
	\sigma_{E_{J}}=\sqrt{K}\,\sigma_{n}\,|\Delta E| .
	\label{eqEJstats}
\end{equation}
Thus $\langle E_{J}\rangle$ scales with the junction area $A$, whereas $\sigma_{E_{J}}$ scales with $\sqrt{A}$, so that the relative spread decreases as $A^{-1/2}$. For a junction with $A=200\times 200$ nm$^{2}$, we have $K\approx 3400$, with $A_{1}=34.17\times 34.17$ \AA$^{2}$, $A_{0}=9.61\times 8.32$ \AA$^{2}$, and $\Delta =0.20$ meV. At the average H content of $2.56$ at.$\%$, the Josephson energy is
\begin{equation}
	\frac{\langle E_{J}\rangle}{h}=10.23\ \text{GHz},\qquad \frac{\sigma_{E_{J}}}{h}=3.6\ \text{MHz},
	\label{eqEJresult}
\end{equation}
corresponding to a relative spread of only $3.5\times 10^{-4}$. Hydrogen contamination therefore shifts the mean Josephson energy by $10.6\%$ relative to that of the pure junction ($E_{J,\text{JJ}}/h=9.26$ GHz), while its contribution to device-to-device variability remains significantly smaller than the variability observed in state-of-the-art fabricated junctions: the relative spread is more than two orders of magnitude smaller than the spread of about $10\%$ caused by interface roughness in junctions of the same area~\cite{ref-Zhu2026-APL}. The H-induced shift of the mean is also small compared with the effect of the barrier thickness. NEGF-DFT calculations for Al/$\alpha$-Al$_2$O$_3$/Al junctions show that the zero-bias conductance decays exponentially with the oxide thickness, by a factor of about $4$--$7$ per added oxide layer ($\approx 2.2$~\AA), and that the interface termination changes the conductance by up to an order of magnitude~\cite{ref-Shan2022}. Applied to the present junction, a barrier one monolayer thicker or thinner would change $E_{J,\text{JJ}}/h$ from $9.26$~GHz to about $1.3$--$2.3$~GHz or $37$--$65$~GHz, respectively, compared with the $11\%$ shift caused by hydrogen. Variations in the barrier thickness and interface termination therefore dominate the variability of $E_{J}$, while hydrogen contamination mainly causes a systematic shift of its mean.

To obtain the dependence on the hydrogen concentration $c$, we assume that the hydrogen uptake mechanism does not depend on $c$: $\alpha$ and $\beta$ in Eq.~(\ref{eq10}) are kept fixed, and $M$ is scaled in proportion to $c$, i.e., $M(c)=M_{0}\,c/c_{0}$, where $c_{0}=2.56$ at.$\%$ and $M_{0}=40$ are the values from our simulations. Then $\langle n\rangle$ is proportional to $c$, and $\sigma_{n}$ follows from the variance of the beta-binomial distribution. As shown in Table~\ref{tabEJ}, the mean $E_{J}$ increases linearly with the hydrogen concentration, while the relative spread remains below $10^{-3}$. This scaling assumption is supported by additional MD simulations with different amounts of H$_2$O, in which the number of H atoms in the oxide increases approximately linearly with the amount of H$_2$O (Appendix~\ref{app:NH2O}).

\begin{table}[htbp]
	\centering
	\caption{Mean $\langle E_{J}\rangle$ and standard deviation $\sigma_{E_{J}}$ of the Josephson energy of a $200\times 200$ nm$^{2}$ junction for different hydrogen contents.}
	\label{tabEJ}
	\begin{tabular}{ccc}
		\hline\hline
		H content (at.$\%$) & $\langle E_{J}\rangle /h$ (GHz) & $\sigma_{E_{J}}/h$ (MHz) \\ \hline
		$1.3$ & $9.75$ & $2.2$ \\
		$2.56$ & $10.23$ & $3.6$ \\
		$5.1$ & $11.20$ & $6.3$ \\
		\hline\hline
	\end{tabular}
\end{table}

\section{Summary}

To summarize, we investigated hydrogen contamination in Al/AlO$_{\text{x}}$%
/Al Josephson junctions using MD simulations with CHGNet and atomistic
quantum transport modeling with NanoDCAL: the MD simulations give the number and binding sites of the H atoms, the transport calculations give the change in transmission caused by one H atom, and a statistical model combines the two into the distribution of $E_{J}$. It is found that the number
of H atoms across different Al/AlO$_{\text{x}}$ tunnel barriers is well
described by a beta-binomial distribution (Fig. \ref{fig02}). Across $400$
Al/AlO$_{\text{x}}$ interfaces, $91.3\%$ of the H atoms form terminal ($\mu_{1}$) or bridging ($\mu_{2}$ and, rarely, $\mu_{3}$) hydroxyl groups bonded to Al, $5.4\%$ H atoms bind to an Al atom through
chemisorbed H$_{\text{2}}$O-like motifs, while the remainder form a variety
of rare configurations (Table \ref{tab01}). Regarding transport properties,
the positively charged H atom forms an interface dipole that raises the oxide bands by $\sim 0.8$ eV relative to the Fermi energy, moving $E_{F}$ toward the valence band (Fig.~\ref{fig04}). Finally, the Josephson energy distribution is obtained as $E_{J}/h=10.23$ GHz with a standard deviation of $3.6$ MHz (Table~\ref{tabEJ}), corresponding to an average H atom content of $2.56$ at.$\%$ in a $200\times 200$ nm$^{2}$ junction. Hydrogen thus shifts the mean Josephson energy by about $11\%$, linearly with the hydrogen content, while the random incorporation of hydrogen gives a relative spread of only $3.5\times 10^{-4}$. Within a fabrication run, the contribution of hydrogen contamination to device-to-device variability is therefore significantly smaller than the variability observed in state-of-the-art fabricated junctions. In contrast, run-to-run differences in the hydrogen environment during oxidation, such as the residual water pressure, are predicted to shift $E_{J}$ systematically. This prediction can be tested experimentally by oxidizing junctions under controlled partial pressures of water.

The present work focuses on the effect of hydrogen impurities on the Josephson energy of Josephson junctions and its device-to-device variability. The H motifs identified in Sec.~III are candidates for H-related TLS, which limit the coherence time of superconducting qubits. Whether a motif acts as a TLS depends on the potential-energy landscape of the H atom, including the barrier height, the asymmetry, and the tunneling splitting between nearby minima. A preliminary analysis of H transfer between neighboring O atoms, presented in Appendix~\ref{app:doublewell}, yields double-well potentials with level spacings of about $90$--$190$~GHz for two candidate sites, well above typical qubit frequencies; a statistical survey over many sites is needed to assess H-related TLS. The motif statistics obtained here, together with this preliminary analysis, provide a starting point for such a systematic study of the potential-energy landscape, which will be reported separately. Such H atoms switching between two configurations would also modulate the local tunneling transmission in time. In the model of Constantin and Yu~\cite{ref-ConstantinYu2007}, two-level fluctuators in the tunnel barrier modify the barrier potential and together produce the $1/f$ critical-current noise observed in Josephson junctions~\cite{ref-VanHarlingen2004}. The present work provides two ingredients of such a model for hydrogen: the change in the transmission coefficient caused by an H atom ($7.2\%$), which sets the scale of the coupling, and the number and bonding motifs of the H atoms (about $7\times 10^{4}$ in a $200\times 200$ nm$^{2}$ junction, mostly in Al-OH and Al-OH-Al motifs). The remaining ingredient, the energy landscape that determines the switching rates, is left for future work. Broadly, the multiscale theoretical framework combining machine-learning MD with NEGF-DFT transport analysis presented here can be extended to investigate other common contaminants such as carbon or nitrogen, in superconducting qubit junctions. We hope to be able to present these investigations in a future opportunity. 

\section*{Data Availability}
The amorphous Al$_2$O$_3$ DFT dataset used to validate CHGNet (Appendix~\ref{app:chgnet}) is publicly available~\cite{ref-PAA2026}. The pretrained CHGNet model is publicly available~\cite{ref-chgnet}. The other data generated in this study are not publicly available. Requests for further information or data should be sent to the authors.

\section*{Acknowledgments}
We thank the Digital Research Alliance of Canada for the computational facilities that made this work possible. H.G. is grateful to Natural Science and Engineering Research Council (NSERC) of Canada for partial financial support. The authors thank Prof. Jared Cole for stimulating discussions on neutron-scattering experiments for characterizing Josephson junctions, and Dr. Wanting Zhang for providing optimized atomic basis sets which we used in transport calculations of this work.

\appendix

\section{Benchmark of CHGNet}
\label{app:chgnet}

\begin{figure}[t]
	\centering
	\includegraphics[width=\columnwidth]{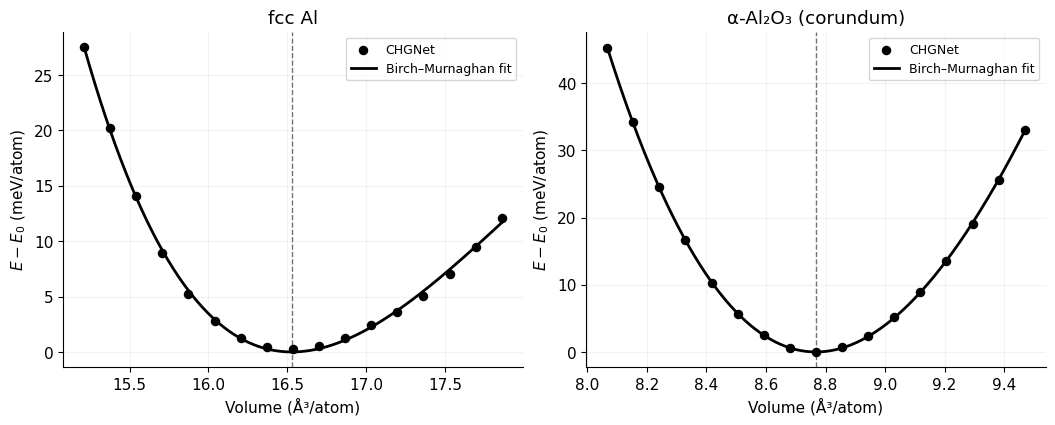}
	\caption{Equations of state of fcc Al and $\alpha$-Al$_2$O$_3$ computed with CHGNet (symbols) and fitted to the third-order Birch--Murnaghan form (lines). The vertical dashed lines mark the equilibrium volumes.}
	\label{figA1}
\end{figure}

\begin{figure}[t]
	\centering
	\includegraphics[width=\columnwidth]{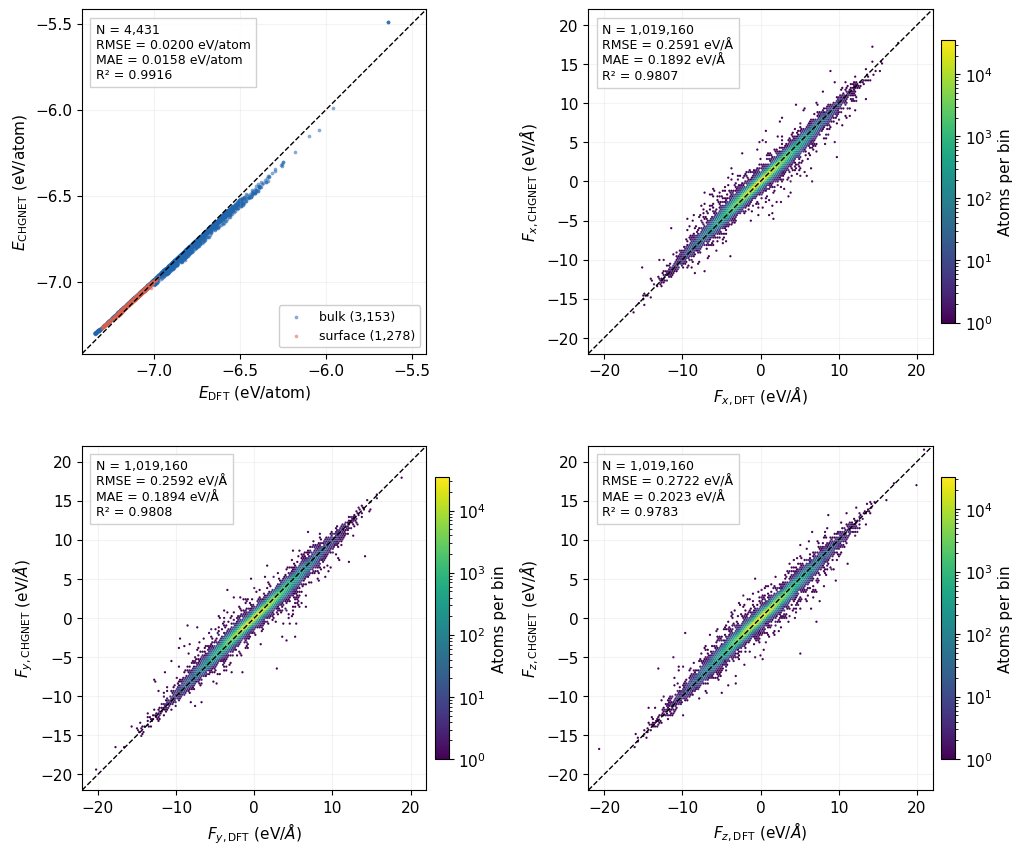}
	\caption{Energies per atom and Cartesian force components predicted by CHGNet versus DFT for the 4{,}431 amorphous Al$_2$O$_3$ configurations of Refs.~\cite{ref-Yu2026-amorphous-alumina,ref-PAA2026}. The CHGNet energies are shifted by a constant offset of $0.44$~eV/atom. The color scale in the force panels gives the number of atoms per bin.}
	\label{figA2}
\end{figure}

\begin{figure}[t]
	\centering
	\includegraphics[width=\columnwidth]{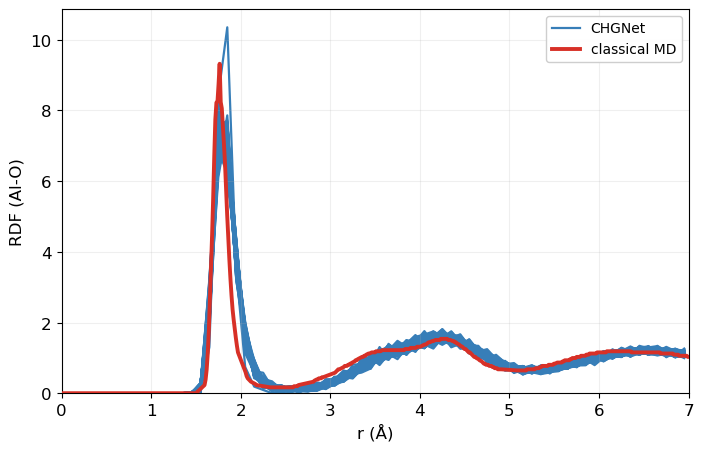}
	\caption{Al--O radial distribution functions of the 400 AlO$_{\text{x}}$ layers generated with CHGNet (blue curves) compared with that of bulk amorphous Al$_2$O$_3$ from classical MD~\cite{ref-Gutierrez2002} (red curve).}
	\label{figA3}
\end{figure}

\begin{table*}[t]
	\centering
	\caption{Lattice parameters and bulk moduli $B_0$ of fcc Al and $\alpha$-Al$_2$O$_3$ obtained with CHGNet, compared with DFT and experiment. The DFT values were obtained with the PBE functional or the local-density approximation (LDA).}
	\label{tabA1}
	\begin{tabular}{llccc}
		\hline\hline
		& & CHGNet & DFT & Experiment \\ \hline
		fcc Al & $a$ (\AA) & $4.044$ & $4.053$ (PBE)~\cite{ref-Schultz2021} & $4.032$ (0~K)~\cite{ref-Schultz2021} \\
		& $B_0$ (GPa) & $55.5$ & $76.2$ (PBE)~\cite{ref-Schultz2021} & $79.4$ (4~K)~\cite{ref-Schultz2021} \\
		$\alpha$-Al$_2$O$_3$ & $a$ (\AA) & $4.812$ & $4.805$ (PBE)~\cite{ref-vanGog2021} & $4.759$ (296~K)~\cite{ref-NIST-SRM1976b} \\
		& $c$ (\AA) & $13.115$ & $13.114$ (PBE)~\cite{ref-vanGog2021} & $12.993$ (296~K)~\cite{ref-NIST-SRM1976b} \\
		& $B_0$ (GPa) & $219.2$ & $248.7$ (LDA)~\cite{ref-Boettger1997} & $252.0$ (0~K)~\cite{ref-Tefft1966} \\
		\hline\hline
	\end{tabular}
\end{table*}

The MD simulations in Sec.~II use the pretrained CHGNet model (checkpoint 0.3.0) without fine-tuning~\cite{ref-chgnet}. Here we assess its accuracy for the Al--O system with three tests.

\textit{Crystalline phases.} The equations of state of fcc Al and $\alpha$-Al$_2$O$_3$ were computed at 17 volumes within $\pm 8\%$ of equilibrium, relaxing the internal coordinates and $c/a$ of $\alpha$-Al$_2$O$_3$ at each volume, and fitted to the third-order Birch--Murnaghan form (Fig.~\ref{figA1}). As shown in Table~\ref{tabA1}, the lattice parameters agree with PBE, the functional of the CHGNet training data, to within $0.3\%$, and with experiment to within $1.1\%$. The bulk moduli are underestimated by $30\%$ for Al and $13\%$ for $\alpha$-Al$_2$O$_3$. CHGNet thus describes the equilibrium bond lengths well but underestimates the compressional stiffness.

\textit{Amorphous Al$_2$O$_3$.} We evaluated CHGNet on the Princeton amorphous alumina dataset~\cite{ref-Yu2026-amorphous-alumina,ref-PAA2026}, which contains 4{,}431 PBE configurations of amorphous Al$_2$O$_3$ (3{,}153 bulk and 1{,}278 surface structures, $1.02\times10^{6}$ atoms in total). After a single constant energy offset of $0.44$~eV/atom, which accounts for the different energy references, CHGNet reproduces the DFT energies with a mean absolute error (MAE) of $16$~meV/atom (root-mean-square error $20$~meV/atom, $R^2=0.99$) and the forces with an MAE of $0.19$~eV/\AA\ (root-mean-square error $0.26$~eV/\AA, $R^2=0.98$), for bulk and surface structures alike (Fig.~\ref{figA2}).

\textit{Structure of the simulated oxide.} Figure~\ref{figA3} compares the Al--O radial distribution function of the 400 AlO$_{\text{x}}$ layers produced by our oxidation simulations with that of bulk amorphous Al$_2$O$_3$ from classical MD~\cite{ref-Gutierrez2002}. The first peak lies at $1.79\pm0.05$~\AA, compared with $1.76$~\AA\ in the reference, and the second shell near $4.2$~\AA\ is reproduced. The broader first peak of our structures is expected, since the simulated oxide is substoichiometric (O/Al $\approx 1.25$), contains hydrogen, and has two interfaces.

In the literature, CHGNet models fine-tuned on system-specific DFT data have been used for molecular dynamics of amorphous lithium fluorophosphates~\cite{ref-Zhong2026}, amorphous alkali oxyhalides~\cite{ref-Binci2026}, and amorphous alloys~\cite{ref-Gong2025}, and CHGNet is among the potentials supported by the atomate2 workflows, including the MPMorph workflow for amorphous structures~\cite{ref-Ganose2025}. 

The quantities extracted from our MD simulations, namely the number of incorporated H atoms, their bonding motifs, and the oxide thickness, are governed by bond formation and local bonding environments. The accuracy of the energies, forces, and bond lengths found above is adequate for this purpose. We do not use the simulations for quantities that require accurate elastic constants or absolute formation energies, and the MD results should be read as statistical trends rather than quantitatively exact numbers.

\section{Dependence on the O$_2$ density}
\label{app:NO2}

\begin{figure}[t]
	\centering
	\includegraphics[width=\columnwidth]{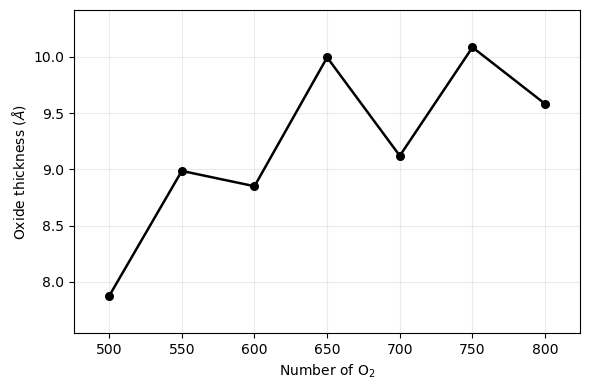}
	\caption{Oxide thickness as a function of the number of O$_2$ molecules in the MD simulations without H$_2$O. Each point is obtained from a single simulation.}
	\label{figB1}
\end{figure}

To examine the sensitivity of the oxide to the O$_2$ density used to accelerate the oxidation (Sec.~II), we performed additional MD simulations without H$_2$O, in which the number of O$_2$ molecules is varied from 500 to 800 in the same simulation cell. All other parameters are the same as in Sec.~II. The oxide thickness is defined as the extent along $z$ of the O atoms bonded to Al (Al--O distance below $2.3$~\AA), between the 2.5th and 97.5th percentiles of their $z$ coordinates.

As shown in Fig.~\ref{figB1}, the oxide thickness increases from $0.79$~nm with 500 O$_2$ molecules to $1.00$~nm with 650 O$_2$ molecules. Above 650 O$_2$ molecules, it saturates and remains between $0.91$ and $1.01$~nm up to 800 O$_2$ molecules. The variation in this range is within the thickness fluctuations of a single simulation. The number of O$_2$ molecules used in the production runs (750) lies in this saturated regime, so the oxide structure is insensitive to the precise O$_2$ density.

\section{Dependence on the amount of H$_2$O}
\label{app:NH2O}

\begin{figure}[t]
	\centering
	\includegraphics[width=\columnwidth]{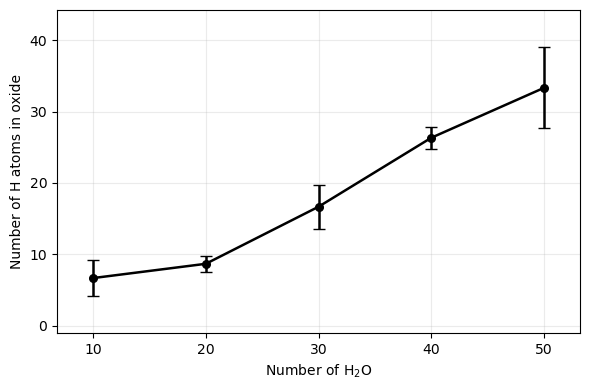}
	\caption{Number of H atoms in the oxide as a function of the number of H$_2$O molecules, with 750 O$_2$ molecules. The symbols and error bars give the mean and standard deviation over three independent samples.}
	\label{figC1}
\end{figure}

To examine how the hydrogen content of the oxide depends on the amount of water, we performed additional MD simulations with 750 O$_2$ molecules and 10, 20, 30, 40, and 50 H$_2$O molecules, using three independent samples with different random initial configurations for each amount of H$_2$O. All other parameters are the same as in Sec.~II. An H atom is counted as incorporated in the oxide if it lies within $1.25$~\AA\ of an O atom that is within $2.5$~\AA\ of an Al atom.

As shown in Fig.~\ref{figC1}, the number of H atoms in the oxide increases monotonically with the number of H$_2$O molecules, from $6.7\pm 2.5$ for 10 H$_2$O molecules to $33.3\pm 5.7$ for 50 H$_2$O molecules, and is approximately proportional to it, with about $0.6$ H atoms incorporated per H$_2$O molecule. This supports the assumption in Sec.~V that the mean number of H atoms scales linearly with the hydrogen supply while the incorporation mechanism remains unchanged. For 30 H$_2$O molecules, the mean of the three samples ($16.7\pm 3.1$) is somewhat lower than, but comparable to, that of the 400 production samples ($21.41\pm 4.62$), considering the small number of samples.

\section{Oxide growth}
\label{app:growth}

\begin{figure}[t]
	\centering
	\includegraphics[width=\columnwidth]{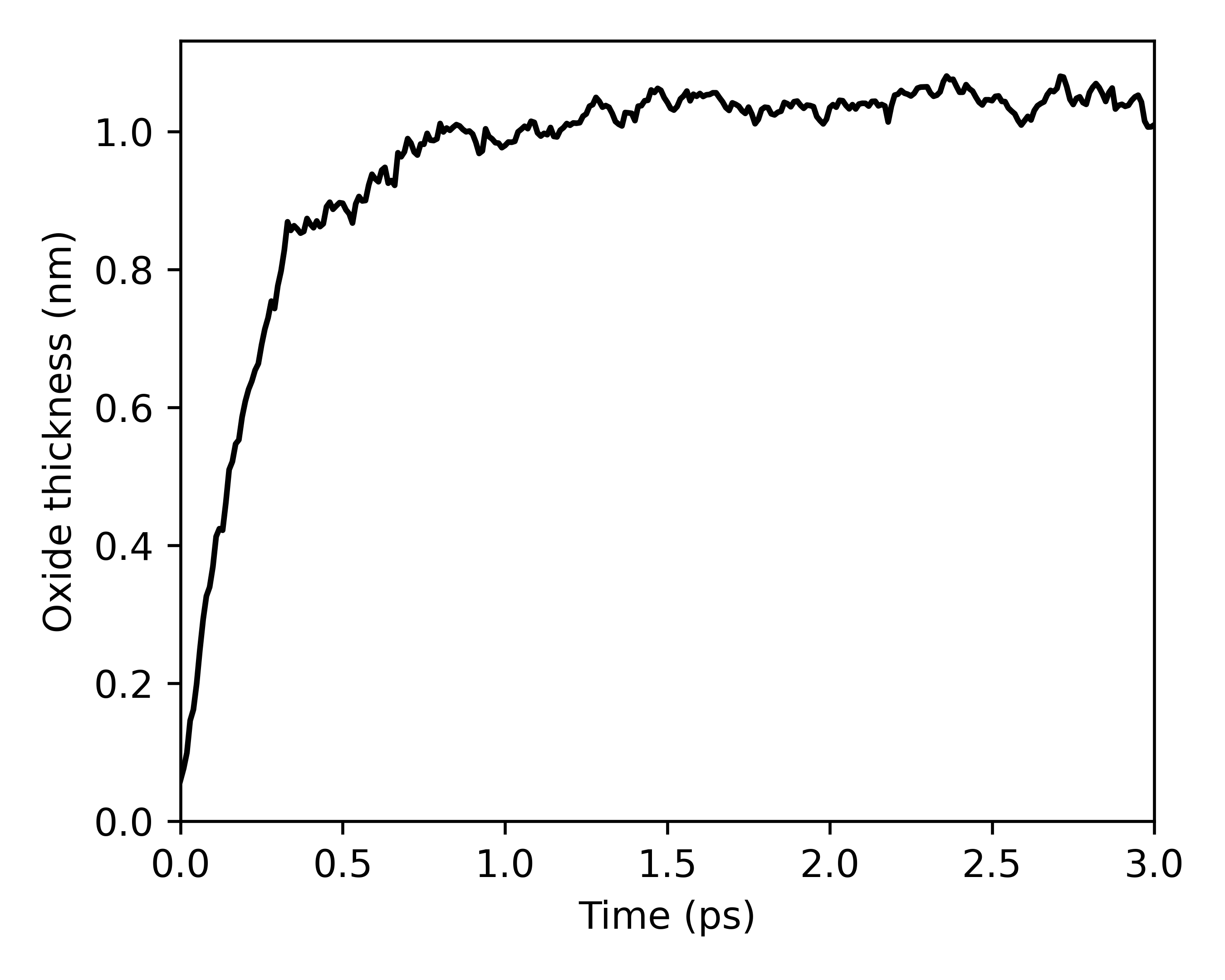}
	\caption{Oxide thickness as a function of simulation time for a production MD simulation with 750 O$_2$ and 30 H$_2$O molecules.}
	\label{figD1}
\end{figure}

Figure~\ref{figD1} shows the oxide thickness as a function of simulation time for one of the production MD simulations, with the thickness defined as in Appendix~\ref{app:NO2}. Oxidation is rapid during the first $\sim 0.3$~ps, when the oxide grows to about $0.8$~nm. It then slows down markedly and saturates at about $1.0$~nm after $\sim 1.5$~ps. About $70\%$ of the O$_2$ molecules remain unreacted in the gas at the end of the simulation, so the saturation is not caused by depletion of the oxidant. Instead, it reflects the self-limiting nature of Al oxidation: once an oxide of about $1$~nm has formed, it blocks further oxidation even at the high O$_2$ density used here. The final thickness is consistent with the $1$--$2$~nm tunnel barriers of experimental Josephson junctions.

\section{Preliminary double-well analysis}
\label{app:doublewell}

\begin{figure}[t]
	\centering
	\includegraphics[width=\columnwidth]{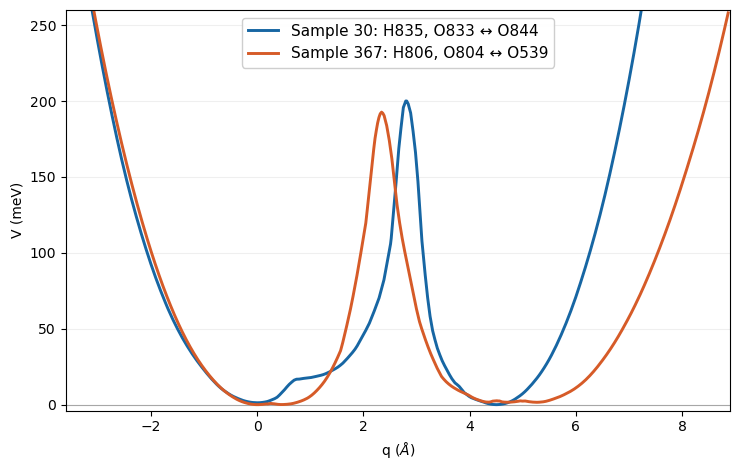}
	\caption{Potential energy $V$ along the minimum-energy path of H transfer between two O atoms for samples 30 and 367, as a function of the mass-weighted collective coordinate $q$. Each curve is measured from its lower minimum. The atom labels give the transferring H atom and the two O atoms.}
	\label{figE1}
\end{figure}

A hydrogen atom can act as a TLS only if it can tunnel between two nearby configurations~\cite{Gordon2014TLS,Holder2013TLS,Muller2019TLSReview}. To illustrate how the H motifs identified in Sec.~III may give rise to such double wells, we carried out a preliminary analysis for a small set of configurations. We selected H atoms bound to O atoms at the oxide surface and searched for nearby O atoms that could serve as alternative binding sites. For each candidate, the two configurations with H bound to either O atom were relaxed, allowing nearby atoms to move while keeping the rest of the structure fixed. The minimum-energy path between them was then determined with the nudged elastic band (NEB) method~\cite{ref-NEB-1998} as implemented in ASE~\cite{ref-ASE-2017}, using CHGNet energies and forces. The motion of all moving atoms is described by the collective coordinate $q$, defined by $dq^{2}=\sum_{i}(m_{i}/m_{\text{H}})\,|d\mathbf{R}_{i}|^{2}$, and the tunneling splitting $\Delta$ and asymmetry $\varepsilon$ were obtained by solving the one-dimensional Schr\"odinger equation for a particle of mass $m_{\text{H}}$ in the potential along the path.

Two H-transfer double wells of the O--H$\cdots$O type proposed by Gordon \textit{et al.}~\cite{Gordon2014TLS} were found in samples 30 and 367 (Fig.~\ref{figE1}). The H atom moves by $1.23$ and $1.04$~\AA\ between the two binding sites, respectively; the larger separations of the minima in Fig.~\ref{figE1} reflect the motion of the surrounding atoms included in $q$. The barriers measured from the lower minimum are $200$ and $193$~meV, and the tunneling splittings are $\Delta/h\approx 0.66$ and $0.38$~MHz. The level spacings $E/h=\sqrt{\varepsilon^{2}+\Delta^{2}}/h\approx 186$ and $94$~GHz are dominated by the asymmetry $\varepsilon$, so these two candidates are far from resonance with qubits operating at $5$--$10$~GHz.

These results are preliminary. The tunneling splitting is sensitive to the path resolution, and the isolated two-level doublet is not retained when a larger region of atoms is allowed to relax. In addition, the one-dimensional model neglects other tunneling paths and transverse vibrations, and the barriers and asymmetries may be affected by the accuracy of the pretrained CHGNet potential. Because the asymmetry varies widely from site to site in an amorphous oxide, a statistical survey over many sites, together with DFT validation or fine-tuning of the potential, is needed before conclusions about H-related TLS can be drawn. A systematic study will be reported separately.

\section{Dependence on the Hubbard parameter $U$}
\label{app:U}

\begin{figure}[t]
	\centering
	\includegraphics[width=\columnwidth]{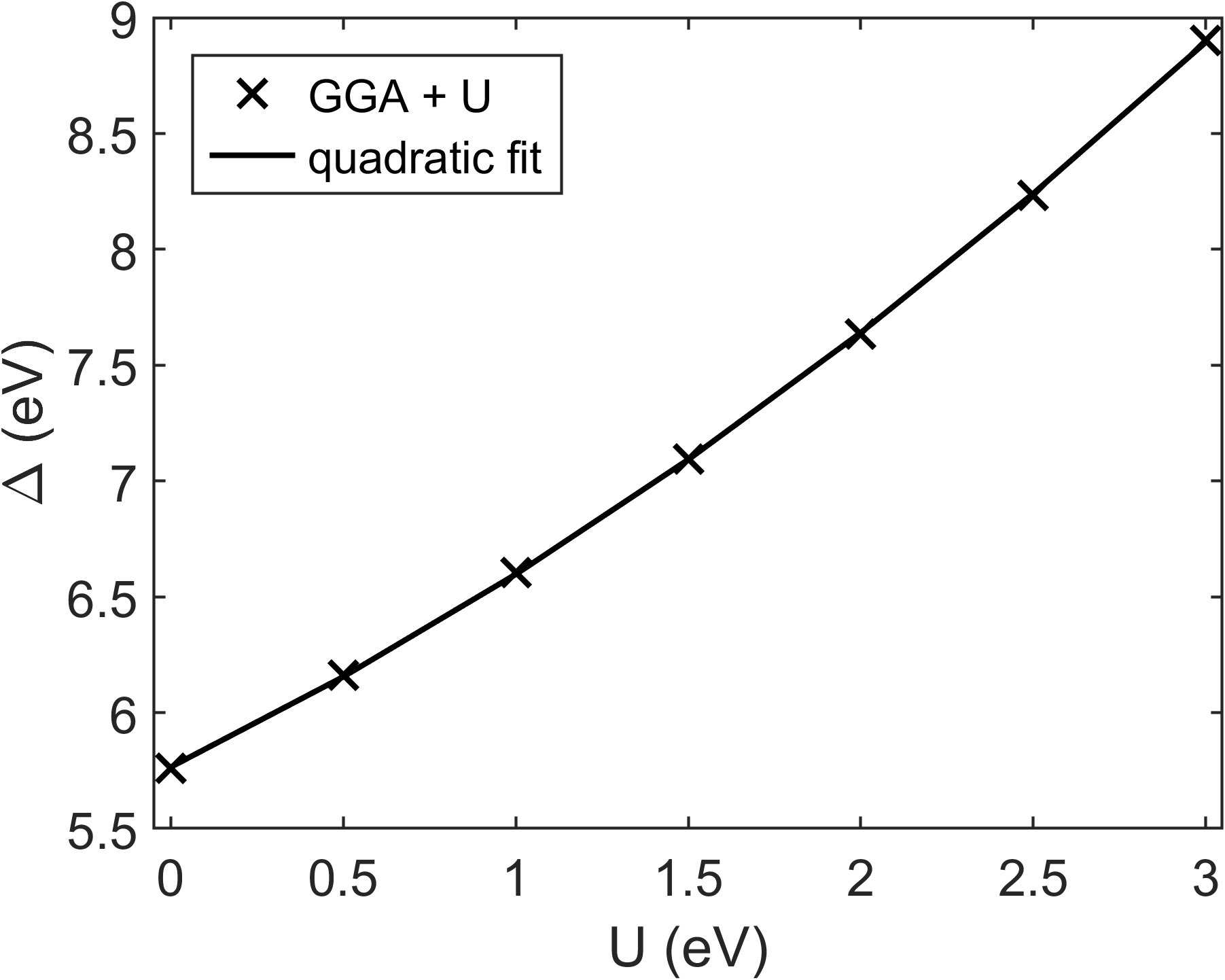}
	\caption{Band gap $\Delta$ of $\alpha$-Al$_2$O$_3$ as a function of $U$, together with a quadratic fit.}
	\label{figS6a}
\end{figure}

\begin{figure}[t]
	\centering
	\includegraphics[width=\columnwidth]{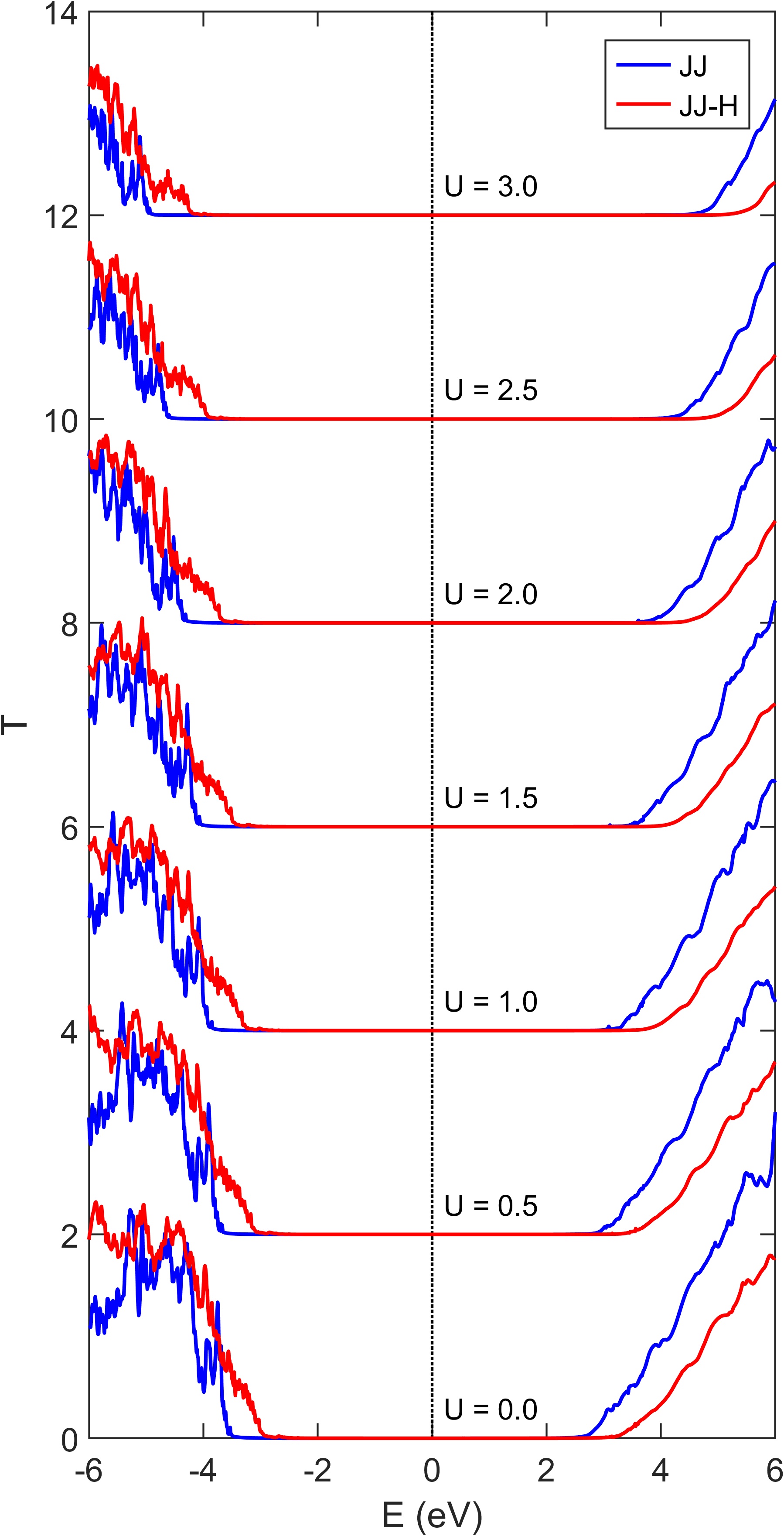}
	\caption{Transmission coefficients $T(E)$ of the JJ (blue curves) and JJ-H (red curves) junctions for different values of $U$. The curves are offset vertically for clarity. The Fermi level is indicated by the vertical dotted line.}
	\label{figS6b}
\end{figure}

This appendix presents the dependence of the band gap of $\alpha$-Al$_2$O$_3$ and of the transmission coefficients of the JJ and JJ-H junctions on the value of $U$, where $U$ is the Hubbard parameter applied to the p orbitals of O atoms to account for electron correlation effects.

Figure~\ref{figS6a} shows the band gap $\Delta$ of $\alpha$-Al$_2$O$_3$ as a function of $U$. As expected, $\Delta$ increases with $U$, from $5.76$~eV at $U=0$ (GGA) to $8.90$~eV at $U=3.0$~eV. The computed band gaps are well described by a quadratic function of $U$, which we use only as an empirical interpolation within the computed range $0\le U\le 3$~eV: its root-mean-square deviation from the data is $6.2$~meV, compared with $92$~meV for a linear fit. From the quadratic fit, the band gap $8.82$~eV obtained with the Heyd--Scuseria--Ernzerhof (HSE) hybrid functional~\cite{ref-HSE-band-gap} is reproduced at $U=2.94$~eV. However, the transport properties are governed by the position of the Al Fermi level relative to the band edges of Al$_2$O$_3$ rather than by the band gap itself. Since $U=1.0$~eV yields the band alignment obtained from internal photoemission measurements~\cite{ref-band-offset-exp} and DFT calculations~\cite{ref-band-offset-thy}, this value is adopted in the transport calculations of the main text.

Figure~\ref{figS6b} shows the transmission coefficient $T(E)$ of the JJ and JJ-H junctions for different values of $U$. As $U$ increases, the transmission gap widens, implying a smaller transmission coefficient at the Fermi level. In addition, for all values of $U$, the presence of the H atom shifts $T(E)$ such that the Fermi level moves toward the valence band edge.
\bibliography{references}

\end{document}